\documentclass[useAMS,usenatbib,fleqn]{mnras}
\usepackage{amsmath}
\usepackage{multirow}
\usepackage{graphicx}
\usepackage{color,soul}
\usepackage{epsfig}
\usepackage{amssymb,amsmath}
\usepackage{aas_macros}
\usepackage{fixltx2e}
\title[SMBH-Galaxy Co-evolution in Romulus]{
Tracing Black Hole and Galaxy Co-evolution in the {\sc Romulus} Simulations
}

\author[Ricarte et al.]{Angelo Ricarte$^1$, Michael Tremmel$^{1,2}$, 
Priyamvada Natarajan$^{1,2}$, Thomas Quinn$^3$ \\
$^1$ Department of Astronomy, Yale University, 52 Hillhouse Avenue, New Haven, CT 06511 \\
$^2$ Department of Physics, Yale University, P.O. Box 208121, New Haven, CT 06520 \\
$^3$ Department of Astronomy, University of Washington, PO Box 351580, Seattle, WA, 98195-1580, USA 
}

\date{\today}

\begin{document}
\pagerange{\pageref{firstpage}--\pageref{lastpage}} \pubyear{2017}
\maketitle

\begin{abstract}
We study the link between supermassive black hole growth and the stellar mass assembly of their host galaxies in the state-of-the-art {\sc Romulus} suite of simulations. The cosmological simulations {\sc Romulus25} and {\sc RomulusC} employ innovative recipes for the seeding, accretion, and dynamics of black holes in the field and cluster environments respectively.  We find that the black hole accretion rate traces the star formation rate among star-forming galaxies.  This result holds for stellar masses between $10^8$ and $10^{12}$ solar masses, with a very weak dependence on host halo mass or redshift.  The inferred relation between accretion rate and star formation rate does not appear to depend on environment, as no difference is seen in the cluster/proto-cluster volume compared to the field.  A model including the star formation rate, the black hole-to-stellar mass ratio, and the cold gas fraction can explain about 70 per cent of all variations in the black hole accretion rate among star forming galaxies.  Finally, bearing in mind the limited volume and resolution of these cosmological simulations, we find no evidence for a connection between black hole growth and galaxy mergers, on any timescale and at any redshift.  Black holes and their galaxies assemble in tandem in these simulations, regardless of the larger-scale intergalactic environment, suggesting that black hole growth simply follows star formation on galactic scales.
\end{abstract}


\section{Introduction}\label{sec:intro}

\vspace{20pt}

Every massive galaxy is believed to host a supermassive black hole (SMBH) at its centre \citep{Kormendy&Richstone1995}.  SMBH masses have been found to correlate with the stellar properties of their hosts, namely the bulge luminosity and velocity dispersion, suggestive of a co-evolutionary assembly history between these components \citep{Magorrian+1998,Haehnelt+1998, Ferrarese&Merritt2000,Gebhardt2000,Kormendy&Ho2013,Reines&Volonteri2015,Saglia+2016}.  SMBHs are thought to assemble their masses primarily through luminous gas accretion \citep{Soltan1982}, which powers active galactic nuclei (AGN) that shine at a wide range of wavelengths.  In the standard galaxy evolution paradigm, galaxies must somehow provide the gas reservoir to fuel these AGN, which then impart feedback energy required to suppress star formation in massive haloes \citep{Springel+2005,Croton+2006,Bower+2006}.  The evolving AGN population is now constrained out to $z \lesssim 6$, allowing us to reconstruct the growth of SMBHs throughout cosmic time \citep{Hopkins+2007,Ueda+2014,Aird+2015,Vito+2018,Ananna+2019}. Cosmological simulations have proven to be useful tools to better understand this evolution, and are now able to broadly reproduce observed relationships between SMBHs and their host galaxies, albeit with a variety of physical models \citep[e.g.][]{DiMatteo+2005,Hirschmann+2014,Schaye+2015,sijacki+2015,Volonteri+2016,Tremmel+2017}.

However, the precise connection between AGN and their host galaxies, exactly what ``triggers'' a SMBH to grow and shine, is not yet fully understood. Clearly, accretion of gas from the inner regions of galaxy is the principal growth mode for most if not all black holes.  To provide this gas, idealized galaxy merger simulations have implicated mergers as one of the drivers of AGN activity \citep[e.g.,][]{Mihos&Hernquist1996,DiMatteo+2005,Capelo+2015}.  Yet tests of this paradigm in more realistic environments, in both cosmological simulations and with observational data, have yielded mixed results.  Many morphology studies of optical or X-ray AGN hosts have failed to find a connection with mergers \citep{Cisternas+2011,Mechtley+2016,Villforth+2017}.  On the other hand, mergers appear to be more important for infrared-selected samples \citep{Treister+2010,Glikman+2015,Kocevski+2015,Fan+2016,Ricci+2017,Donley+2018}, which are more complete to obscured accretion \citep[e.g.,][]{Hickox&Alexander2018}.  Targeted ALMA observations have also revealed statistically significant overabundances of companions among the most luminous quasars at $z\sim 5$ \citep{Trakhtenbrot+2017,Trakhtenbrot+2018}.  One possibility is that mergers only trigger the most luminous AGN, while secular processes power ``typical'' moderate-luminosity Seyferts \citep{Treister+2012,Hickox+2014}.  This question of the role that mergers play in the mass assembly history of black holes therefore appears to depend on the mass, luminosity, and selection of these sources.

Many cosmological simulations similarly point towards merger-independent channels of SMBH growth.  Mergers are associated with AGN in the {\sc EAGLE} cosmological simulations, with a more compelling link at low-redshift, perhaps because gas is less readily available at late times \citep{McAlpine+2018}.  Yet analyzing the {\sc Magneticum} pathfinder simulations, \citet{Steinborn+2018} find that mergers only play a minor role in triggering AGN.  In these simulations, the association of the most luminous AGN with mergers can to some extent be explained by the fact that more massive galaxies are inherently more likely to be in merging systems. Similarly, only 35 per cent of SMBH growth is attributed to mergers in the {\sc Horizon-AGN} simulations, the rest originating from unknown secular processes \citep{Martin+2018}. Using a series of zoom-in simulations, \citet{Pontzen+2017} show that galaxy mergers are important for igniting AGN-driven outflows and quenching star formation in massive, high-redshift galaxies, but not for triggering large amounts of SMBH growth. These results of course hinge on the resolution of these simulations, as well as the sub-grid prescriptions for SMBH growth and star formation.  Thus far, it is plausible that mergers, while important for shaping the properties of host galaxies, might not play a significant role for the SMBHs that they harbour.

More broadly, there is currently no consensus on what spatial scales are relevant for SMBH growth.  Do AGN care about the intergalactic environment, only the gas within its sphere of influence, or some scales in between?  At $z\sim 6$, the number density of luminous quasars detectable by SDSS corresponds to a typical host halo mass of $\sim 10^{13} \rm{M}_\odot$ at that epoch, implying that only highly biased $5\sigma$ peaks in the density field could host the fastest growing SMBHs \citep{Fan+2003}.  AGN clustering studies at a wide range of redshifts and luminosities also yield approximately $10^{13} \ \mathrm{M}_\odot$ as the characteristic host halo mass \citep[see e.g.,][for a review]{Cappelluti+2012}.  Consequently, it was expected that we should find galaxy overdensities in the vicinity of the most luminous quasars.  However, no correlation has been found between luminous quasars and protocluster regions at $z\sim 3.8$ \citep{Uchiyama+2018}.  A similar negative result has been found for millimeter source overdensities at $6<z<7$, albeit within ALMA's small field of view \citep{Champagne+2018}.  Using a large cosmological simulation, \citet{DiMatteo+2017} found that the most massive SMBHs were found not in the largest overdensities, but rather in areas with the lowest tidal fields.  Further complicating this picture, feedback from quasars may be capable of inhibiting star formation in nearby haloes, potentially hiding the overdensities in which they reside \citep{Habouzit+2018}.  How the intergalactic environment influences SMBH growth, if at all, is a rapidly evolving field of research.  

One aspect of SMBH astrophysics that complicates all of these studies is that AGN are variable on every timescale \citep{Hickox+2014,Sartori+2018}, and in particular on shorter timescales than that of star formation.  Circumstantial evidence exists for AGN ``flickering'' on timescales of $\sim 10^{4-5}$ yr if optically-elusive X-ray AGN are interpreted as AGN which have not yet had time to photoionize their host galaxies \citep{Schawinski+2015}.  This argument is supported by hydrodynamical simulations with very fine time resolution \citep{Novak+2011} as well as analytic considerations based on the maximum sizes of disks which do not fragment under their self-gravity \citep{King&Nixon2015}.  Recently, several ``changing look'' AGN have also been identified, which have been observed to switch between unobscured (type I) and obscured (type II) accretion states along with similar changes in their levels of continuum emission on $\lesssim$ 10 year timescales \citep[e.g.,][]{LaMassa+2015,MacLeod+2016}.  AGN variability may wash out true correlations that exist on longer timescales, and it is important to better understand this behavior. 

In this paper, we follow and study the demographics and assembly of SMBHs in the {\sc Romulus} simulations, which implement state-of-the-art recipes for SMBH seeding, accretion, and dynamics.  {\sc Romulus25} is a uniform 25 Mpc-per-side volume \citep{Tremmel+2017} that represents the field environment, while {\sc RomulusC} is a zoom-in on a low-mass cluster of $1.5 \times 10^{14} \ \rm{M}_\odot$ \citep{Tremmel+2019}.  Both are run to $z \approx 0$.  This suite of simulations therefore brackets the range of environments in which black holes and galaxies can grow.  Cosmological simulations such as the {\sc Romulus} suite allow us to study SMBH-galaxy co-evolution in realistic large-scale environments.  Although computational limitations restrict our ability to directly resolve star formation and SMBH accretion processes, these complex processes are approximated using innovative and physically-motivated ``sub-grid'' recipes.  One unique advantage of the {\sc Romulus} suite is the fact that the same sub-grid recipes have been implemented for both sets of environments---the field and a cluster.

Here, we explore the relations linking galactic and intergalactic properties to the accretion rates of their SMBHs.  In particular, a causal explanation of the local correlation between SMBH and host stellar masses implies that there must exist some timescale over which the black hole accretion rate and the star formation rate trace each other.  In the past decade, many observational studies have been performed to test this hypothesis, with mixed results \citep[e.g.,][]{Mullaney+2012,Stanley+2015,Aird+2019}.  We demonstrate in this work that the SMBH accretion rate follows the star formation rate well when smoothed to timescales of $\sim 300$ Myr, and that the star formation rate, along with two other parameters, can explain up to 68 per cent of the variations in the SMBH growth rate among star forming galaxies.  Overall, these simulations suggest that SMBHs and their hosts grow in lockstep, but with different variability timescales driven by the stocasticity in both the physics of star-formation and black hole accretion.  Mergers play no noticeable role for the AGN in {\sc Romulus}, but we note that co-evolution might well proceed differently for rare objects that would not appear in its volume:  the most massive SMBHs, luminous SDSS quasars, and transient objects such as ultra-luminous infrared galaxies (ULIRGs).

Our paper is organized as follows: in \S\ref{sec:simulations}, we describe the pertinent details of the {\sc Romulus}  simulations.  In \S\ref{sec:results}, we report the results of our analysis, including derived local relations (\S\ref{sec:local_relations}), the SMBH accretion density (\S\ref{sec:accretion_density}), the connection between AGN and star formation rates (\S\ref{sec:agnms}) followed by a statistical search for the most fundamental relation for SMBH growth (\S\ref{sec:statistical_search}), Eddington ratio distributions (\S\ref{sec:erdfs}), and the potential link between AGN and galaxy mergers (\S\ref{sec:triggers}).  We discuss our findings in the context of other recent work in \S\ref{sec:discussion}, and summarize our key findings in \S\ref{sec:conclusion}.

\section{The {\sc Romulus} Simulations}\label{sec:simulations}

In this paper, we examine SMBH assembly in {\sc Romulus25}, a uniform 25 Mpc-per-side box, and {\sc RomulusC}, a zoom-in simulation of a $10^{14} \ \mathrm{M}_\odot$ cluster.  At $z=0.05$, the latest available redshift slice, {\sc Romulus25} contains 5 haloes of at least $10^{13} \ \mathrm{M}_\odot$ and 39 haloes of at least $10^{12} \ \mathrm{M}_\odot$.  These are some of the highest resolution cosmological simulations that exist with comparable volume.  SMBHs are treated in a more physical manner than in many previous approaches:  they are seeded based on local gas properties, Bondi accretion is modified to account for angular momentum support, feedback is imparted via a single thermal mode, and dynamics are corrected to account for unresolved dynamical friction.  All free parameters are set using a novel optimization technique to grade a small set of zoom-in simulations against empirical relations. 
Below, we summarize the pertinent aspects of the {\sc Romulus} simulations, and more details on the simulation methodology can be found in \citet{Tremmel+2017,Tremmel+2019}.

\subsection{Numerics}\label{sec:numerics}

The {\sc Romulus} simulations were performed with the Tree + Smoothed Particle Hydrodynamics (SPH) code {\sc ChaNGa} \citep{Menon+2015}.  A standard $\Lambda$CDM cosmology was assumed with the Planck cosmological parameters  \citep{Planck2016}.  Dark matter and gas particles have masses of $3.39 \times 10^5 \ \rm{M}_\odot$ and $2.12 \times 10^5 \ \rm{M}_\odot$ respectively.  Gravity is resolved with a Plummer-equivalent softening length of 250 pc, while hydrodynamics are evaluated with a resolution of 70 pc.  {\sc ChaNGa} includes an updated SPH implementation, allowing for more the accurate modelling of shearing flows with Kelvin-Helmholtz instabilities \citep{Menon+2015,Governato+2015,Wadsley+2017}. The simulation also includes an updated implementation of turbulent diffusion, important for correctly modelling gas thermodynamics and metal distribution \citep{Wadsley+2008,Shen+2010,Wadsley+2017}.

\subsection{Star Formation and Cooling}\label{sec:starFormation}

Star formation is modelled with standard recipes, whereby gas particles with temperatures below $10^4 \mathrm{K}$ and above densities of $0.2\  m_\mathrm{p}/\mathrm{cc}$ may form stars with a characteristic timescale of $10^6$ years and an efficiency of 15 per cent, assuming a Kroupa IMF \citep{Kroupa2001} and with supernova feedback proceeding via the ``blastwave'' implementation \citep{Stinson+2006}.  Cooling in low temperature ($<10^4$ K) gas is regulated by the metal abundance \citep{Guedes+2011} as well as thermal and metal diffusion \citep{Shen+2010,Governato+2015}, but high temperature metal-line cooling is not included \citep[see][and references therein for further discussion and details]{Tremmel+2019}. Molecular hydrogen abundance and cooling is not followed in the simulation, as the resolution is not high enough to properly resolve the multi-phase interstellar medium at that level of detail. 

\subsection{SMBH Seeding}\label{sec:seeding}

The physics of seeding, the abundance of seeds and their initial mass function is one of the key unsolved problems in black hole physics today  \citep[see][for a recent detailed discussion]{Ricarte&Natarajan2018b}.  Leveraging the resolution of the {\sc Romulus} simulations, SMBHs are seeded using recipes based on the local gas properties, rather than simply imposing a halo mass threshold as is often implemented \citep[e.g.,][]{Springel+2005,Vogelsberger+2013,Schaye+2015}.  A gas particle selected to form a star will instead form a SMBH if all of the following criteria hold true:
\begin{itemize}
    \item Its metallicity is less than $3 \times 10^{-4} \ Z_\odot$.
    \item Its mass density is greater than $3 \ m_p/\mathrm{cc}$.
    \item Its temperature is between $9500 \ \mathrm{K}$ and $10000 \ \mathrm{K}$.
\end{itemize}

\noindent These criteria ensure that SMBHs form in regions that are collapsing quickly, but also cooling slowly, limiting seed formation to high-density peaks in the early universe with high Jeans masses.  Seeds in the {\sc Romulus} suite are initialized with masses of $10^6 \ \rm{M}_\odot$, immediately accreting from nearby gas particles if necessary, in order to conserve mass.  Although this seed mass is somewhat high even for optimistic direct-collapse scenarios \citep{Lodato&Natarajan2007}, it is important that seed masses be significantly greater than those of gas or dark matter particles in order to resolve dynamical friction and avoid spurious scattering events.  Note that there are no limitations on either the number of seeds that can form in a single halo or how close together seeds can form.  Consequently, some haloes can form multiple seeds that rapidly merge, resulting in ``effective'' seed masses of a few times $10^6 \ \rm{M}_\odot$.

\subsection{SMBH Accretion}\label{sec:accretion}

SMBH accretion is estimated via a modified Bondi-Hoyle prescription \citep{Bondi&Hoyle1944}, corrected to account for the rotational support of surrounding gas on resolved scales \citep[see][for more details]{Tremmel+2017}. Such rotational support has been shown to be important for SMBH growth and feedback \citep{Hopkins+Quataert2010, Rosas-Guevara+2015, Tremmel+2017}. A SMBH's accretion rate is taken to be the minimum of the Eddington accretion rate and one of the following:

\begin{equation}
    \dot{M}_\bullet = \left( \frac{n}{n_*} \right)^\beta \begin{cases}
    \frac{\upi(GM_\bullet)^2\rho}{(v^2_\mathrm{bulk} + c^2_s)^{3/2}} & \text{if $v_\mathrm{bulk} > v_\theta$} \\ 
    \frac{\upi(GM_\bullet)^2\rho c_s}{(v^2_\theta + c^2_s)^{2}} & \text{if $v_\mathrm{bulk} < v_\theta$}, \\ 
    \end{cases}
\end{equation}

\noindent where $G$ is the gravitational constant, $n_*$ is the star formation threshold number density, $v_\mathrm{bulk}$ is the bulk velocity of the gas, $v_\theta$ is its rotational velocity, $c_s$ is its sound speed, $\rho$ is its mass density, and $\beta$ is a free parameter set to 2 based on the optimization procedure described in \S\ref{sec:optimization}.  The prefactor $(n/n_*)^\beta$ is a boost factor commonly employed to correct for underestimates of the gas density and temperature due to the limited resolution of the simulation \citep{Booth&Schaye2009}.

The {\sc Romulus} simulations only employ one mode of AGN feedback, whereas other simulations sometimes implement two distinguishable modes --- ``quasar/radiative'' and ``radio/mechanical'' modes, for example. In {\sc Romulus}, if some mass $\dot{M}_\bullet \Delta t$ is accreted by a SMBH, then an amount of energy $E = \epsilon_r \epsilon_f \dot{M}_\bullet c^2 \Delta t$ is injected isotropically into the nearest 32 gas particles.  Here, $\epsilon_r$ is the radiative efficiency of the accretion disk (assumed to be $0.1$), $\epsilon_f$ is the feedback coupling efficiency (set to $0.02$), and $c$ is the speed of light.  In order to avoid numerical overcooling due to limited mass and time resolution, gas that receives AGN feedback energy is prevented from cooling for the duration of the SMBH's timestep.  This is meant to approximate a continuous transfer of energy  throughout the SMBH's timestep.  In order to avoid long cooling shutoff times, and to more continuously sample the interaction between SMBHs and nearby gas, SMBHs are placed on the smallest global timestep of the simulation, typically $10^4-10^5$ years.  With both the brief cooling shutoff and the short timesteps, {\sc Romulus} is able to avoid gas overcooling, which often artificially suppresses the effects of thermal feedback models.  As a result, SMBHs in {\sc Romulus} are able to drive powerful outflows that successfully regulate and sometimes quench star formation in massive galaxies \citep{Pontzen+2017,Tremmel+2019}, as well as enrich the circumgalactic medium \citep{sanchez+2018}.

\subsection{SMBH Dynamics}\label{sec:dynamics}

New techniques are employed to accurately estimate the dynamical friction of surrounding matter onto SMBHs in these simulations \citep{Tremmel+2015}.  Unless a correction factor is added, the dynamical friction force onto SMBHs is underestimated due to gravitational softening.  This correction factor is obtained by integrating the \citet{Chandrasekhar1943} formula within the gravitational softening length $\epsilon_g$, resulting in

\begin{equation}
    \boldsymbol{a}_\mathrm{DF} = -4 \upi G^2 M_\bullet \rho(<v_\bullet) \ln \Lambda \frac{\boldsymbol{v}_\bullet}{v_\bullet^3},
\end{equation}

\noindent where $\rho(<v_\bullet)$ is mass density of particles moving slower than the black hole, and $\ln \Lambda = \ln (b_\mathrm{max}/b_\mathrm{min})$, where $b_\mathrm{max}$ is set to $\epsilon_g$ and $b_\mathrm{min}$ is set to the $90^\circ$ deflection radius.  

\subsection{Halo Finding and Data Analysis}\label{sec:halo_finding}

Halo finding is performed with the {\sc Amiga} Halo Finder \citep{Knollmann&Knebe2009}, key properties are calculated using {\sc Pynbody} \citep{Pontzen+2013}, and these are organized into a {\sc TANGOS} database \citep{Pontzen&Tremmel2018}.  Note that {\sc RomulusC} is a zoom-in simulation containing a high-resolution region within a low-resolution dark matter-only region.  Due to the peculiarities of this technique, some haloes can be ``contaminated'' by low-resolution particles.  We therefore exclude from our analysis any haloes for which more than 5 per cent of the dark matter particles originate from the low-resolution region.  

While fitting the stellar-to-halo mass relation, \citet{Munshi+2013} determine that hydrodynamical simulations systematically overestimate stellar masses and underestimate halo masses if these are derived from photometry and a critical density threshold respectively.  Stellar masses derived by simply summing star particles are systematically higher than masses estimated from single-band observations even when realistic surface brightness limitations are taken into account.  Meanwhile, the addition of baryons causes virial radii defined by an over-density threshold to decrease, shrinking halo masses relative to dark matter only simulations.  Throughout this paper, we multiply all stellar masses by $0.6$, and divide all halo masses by $0.8$ in order to account for these two countervailing effects on the halo mass and stellar mass \citep[see Figure 5 of][]{Munshi+2013}.  We include in our analysis any halo which contains at least $10^4$ dark matter particles and that has a stellar mass of at least $10^8 \ \mathrm{M}_\odot$ (post-correction), ensuring that the objects that we study are well-resolved.

\subsection{Systematic Parameter Optimization}\label{sec:optimization}

One unique feature of the {\sc Romulus} simulations is the systematic method by which optimal parameters for sub-grid recipes were honed.  Four different sets of zoom-in simulations were performed with halo masses ranging from $10^{10.5}$ to $10^{12} \ \mathrm{M}_\odot$, with dozens of different sub-grid parameter realizations.  Outputs were then graded based on a combination of empirically measured $z=0$ scaling relations:  the stellar to halo mass relation \citep{Moster+2013}, the HI gas fraction as a function of stellar mass \citep{Cannon+2011,Haynes+2011}, the galaxy specific angular momentum as a function of stellar mass \citep{Obreschkow&Glazebrook2014}, and the SMBH to stellar mass relation \citep{Schramm&Silverman2013}.  The first set of simulations was run varying only star formation parameters, which were then fixed for the subsequent set which optimized SMBH accretion parameters.  

Note that this parameter search was confined to halo masses below which AGN are thought to be the dominant source of feedback energy. Results for halos more massive than $10^{12}$ M$_{\odot}$ were not directly optimized for, and are therefore purely predictions of the simulation as they are not part of the calibration.  Furthermore, only $z=0$ properties were used to anchor and constrain the simulation, so the evolution of galaxies and SMBHs with time is also consequently a prediction of the simulation.

\section{Results}\label{sec:results}

\subsection{Galaxies in the Field and Cluster}

\begin{figure*}
   \centering
   \includegraphics[width=\textwidth]{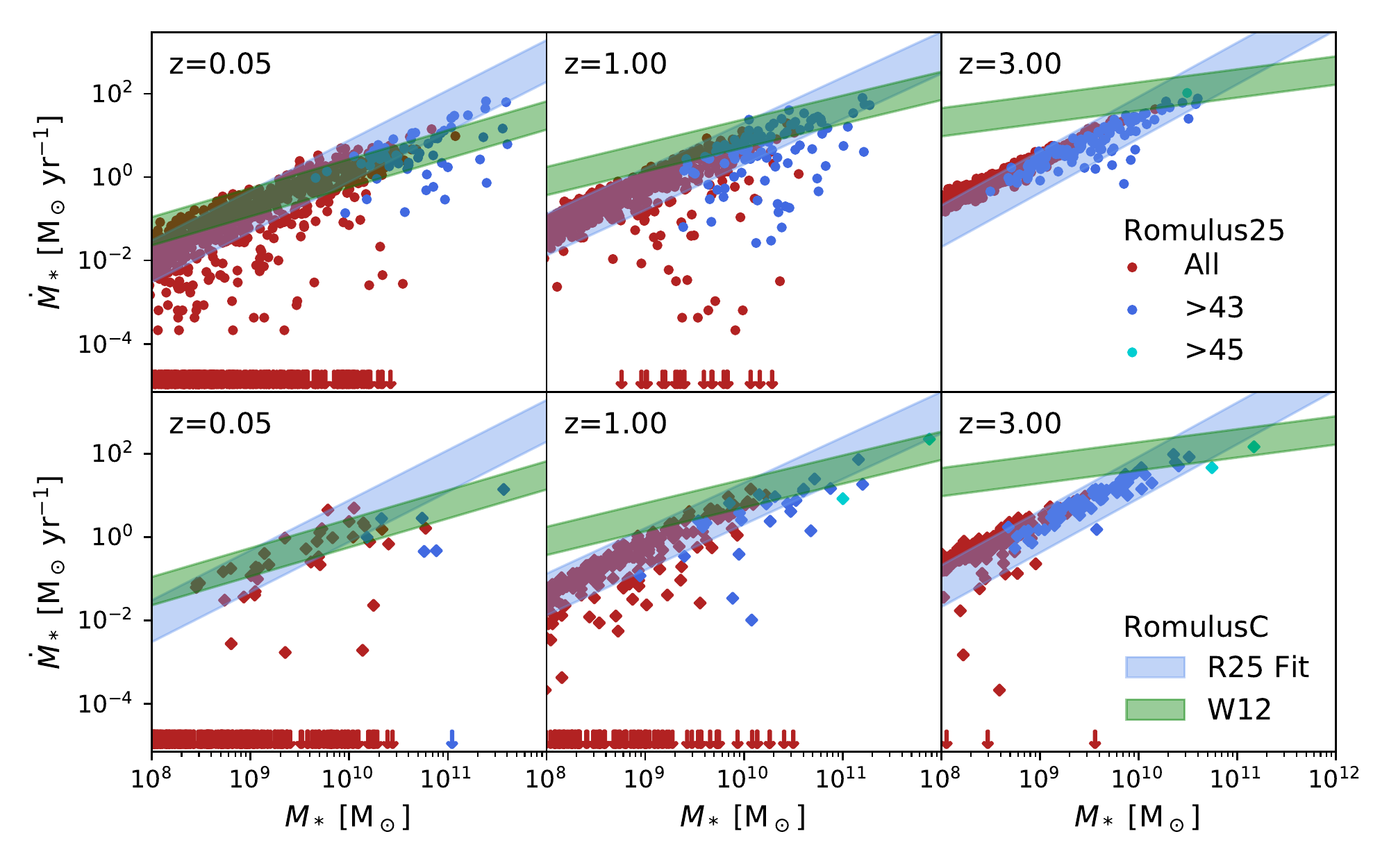}
   \caption{Star formation rates and stellar masses for {\sc Romulus25} and {\sc RomulusC} at redshifts 0.05, 1, and 3.  To illustrate how these galaxies might be selected by a flux-limited AGN survey, points are colour-coded according to the bolometric luminosities of their most massive SMBHs that are hosted in these galaxies.  Both the black hole accretion rates and star formation rates are averaged over 300 Myr, and non-unity duty cycles over this time period would affect selection.  The blue region corresponds to the star forming sequence inferred from the {\sc Romulus25} data points, while the green region corresponds to the empirical relation determined by \citet{Whitaker+2012}.}
   \label{fig:sfms}
\end{figure*}

To set the context of this work, we first begin by examining the galaxy populations of {\sc Romulus25} and {\sc RomulusC}.  In Figure \ref{fig:sfms}, we plot star formation rates (SFRs) versus stellar masses of both simulations for redshifts $z \in \{0.05, 1, 3\}$.  In order to illustrate how these galaxies might be selected in a flux-limited AGN survey, points are colour-coded according to the bolometric luminosities of their most massive\footnote{Throughout this work, we select the most massive SMBH to represent the SMBH/AGN of each host galaxy (and exclude any secondary less massive black holes that might exist in the same halo).  Checking at $z=0.05$, the most massive, most luminous, and most central SMBHs are identical 95.5\% of the time.} SMBHs.  Divisions between tiers are set at $10^{43}$ and $10^{45}$ $\mathrm{erg} \; \mathrm{s}^{-1}$.  Both the SFR and black hole accretion rate (BHAR) are averaged over 300 Myr.  Note that variability can increase the probability that an AGN passes a given flux limit, while also decreasing its duty cycle to conserve mass.  An AGN shining on average at $10^{43} \ \mathrm{erg} \; \mathrm{s}^{-1}$ can be interpreted as an AGN shining at $10^{44} \ \mathrm{erg} \; \mathrm{s}^{-1}$ one tenth of the time, for example.

The green band corresponds to the observed star formation sequence from combined UV and IR SFRs \citep{Whitaker+2012}.  These observations report a shallow, mildly-evolving slope of $\approx 0.6$ and a constant scatter of 0.34 dex.\footnote{For the last panel, equations 1-3 in \citet{Whitaker+2012} have been extrapolated slightly from $z=2.5$ to $z=3$.}  The blue band corresponds to the best fit star forming main sequence derived from central, isolated galaxies in {\sc Romulus25} \citep{Tremmel+2017}. We follow a similar procedure to the observations \citep[e.g.,][]{Bluck+2016} and fit the median values of the star formation rate within 0.1 dex bins of stellar mass between $10^8$ and $10^{10}$ M$_{\odot}$.  Note that the {\sc Romulus} simulations appear to underestimate the star-formation in low-mass galaxies at high-redshift.  However, \citet{Whitaker+2012} is only complete to stellar masses above $\gtrsim 10^{10} \ \mathrm{M}_\odot$ for $z\gtrsim 1$, where the two bands do overlap.  In the work that follows, a galaxy's proximity to the main sequence is computed relative to the fits from the {\sc Romulus} simulations rather than to the observed data, for internal consistency.

As one might expect, many more of the $z=0.05$ galaxies are quenched in the overdense environment of {\sc RomulusC} than in {\sc Romulus25}.  In particular {\sc RomulusC} exhibits a quenched fraction of 80-90 per cent at low masses, compared to 10 per cent in {\sc Romulus25} \citep[see also Figure 14 in][]{Tremmel+2019}. However, among those that are not quenched, there is no indication that their SFRs are different from those of {\sc Romulus25}.  That is, a star-forming galaxy in the cluster or proto-cluster appears to be no different from a star-forming galaxy in the field.  Our colour-coding reveals that the resolution of the {\sc Romulus} simulations allow us to probe much lower galaxy masses than those typically accessible to flux-limited surveys.  X-ray or quasar surveys can typically only reveal the most massive subset of the population we analyze, especially at high-redshift.  It is important to point out that the work we present applies more to typical galaxies than to luminous $>10^{45} \ \mathrm{erg} \; \mathrm{s}^{-1}$ quasars, rare-objects for which {\sc Romulus25} lacks the volume for exploration.

\subsection{Local SMBH-Galaxy Relations}\label{sec:local_relations}

\begin{figure*}
   \centering
   \includegraphics[width=\textwidth]{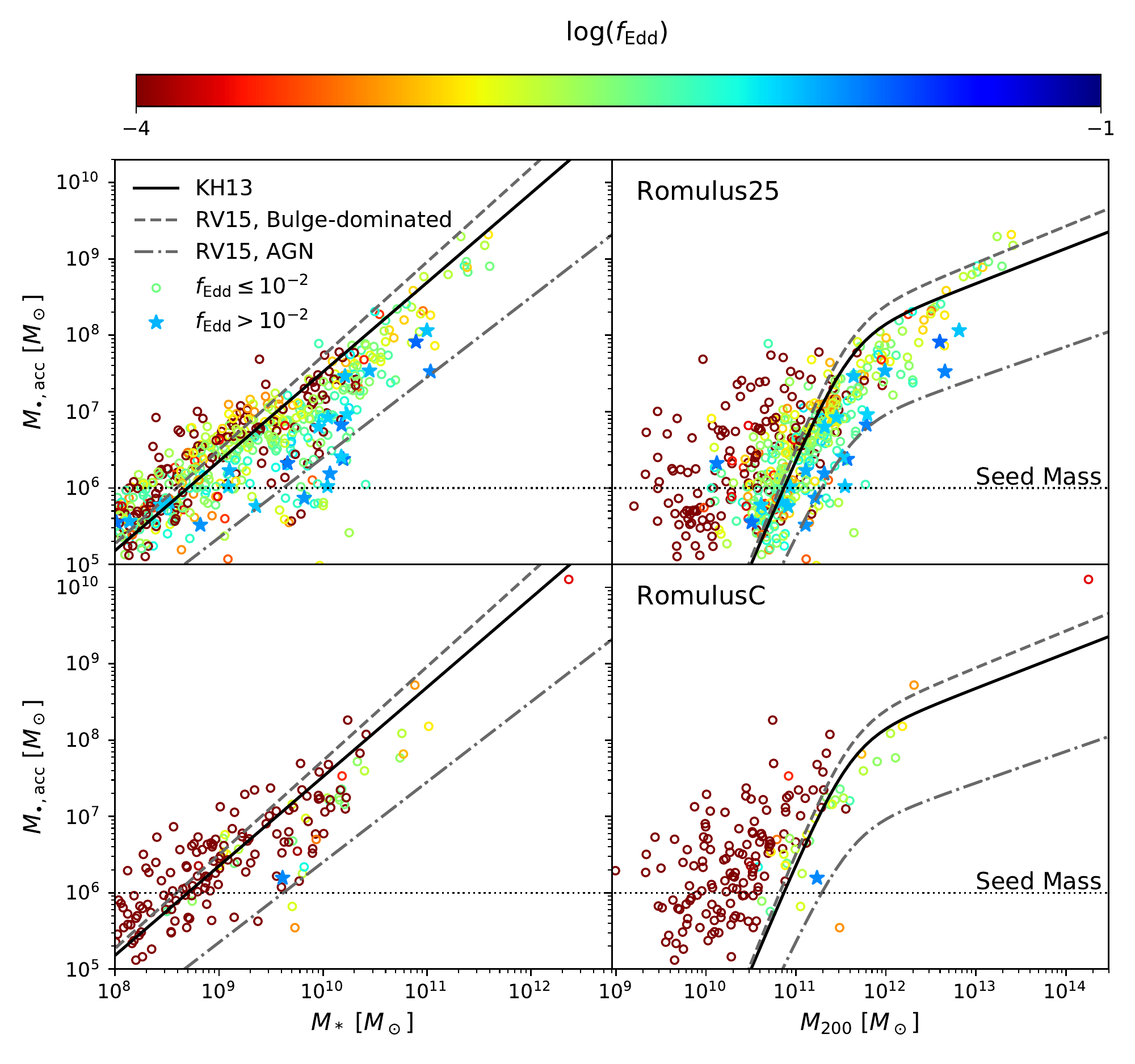}
   \caption{SMBH mass versus host stellar mass ({\it left}) and host virial mass ({\it right}) in the {\sc Romulus} simulations.  Only accreted mass is included, excluding the seed mass contribution.  Relationships from \citet{Kormendy&Ho2013} and \citet{Reines&Volonteri2015} are shown for comparison.  The {\sc Romulus} simulations agree with the relationships observed for bulge-dominated galaxies as calibrated, but with significant scatter at low-masses.  Tidal stripping moves the {\sc RomulusC} data points to the left relative to {\sc Romulus25} in the right panel, but its effects are weak on the host's stellar mass.  Points are colour-coded according to Eddington ratio, revealing a possible explanation for the different relations observed for bulge-dominated galaxies and BLAGN.  If a broad-line region only appears above a certain Eddington ratio threshold, then at fixed accretion rate, lower mass SMBHs are more likely to be BLAGN.}
   \label{fig:mass_relations}
\end{figure*}

We now examine the derived $z=0$ relations between SMBH mass and the host galaxy stellar mass in the {\sc Romulus} suite.  These local relations serve as a boundary condition for determining the assembly history of SMBHs.  There exist well-established observed relations between SMBH masses and the stellar contents of their hosts.  SMBH mass has been shown to correlate with an intrinsic scatter of $\sim 0.3-0.5$ dex with both the luminosity and velocity dispersion of the galactic bulge \citep{Beifiori+2012,Kormendy&Ho2013,vandenBosch2016,Saglia+2016}.  \citet{Reines&Volonteri2015} investigate the relationship between SMBH mass and {\it total} stellar mass, and find that AGN are offset below the relation for bulge-dominated galaxies by over an order of magnitude.  \citet{Shankar+2016} argue that the offset AGN relation may in fact be closer to the true one, and the relation defined by quiescents is significantly biased towards galaxies for which the SMBH sphere of influence can be resolved.  These correlations are important for gaining insight into both fuelling and feedback as a function of host mass.  They are currently adopted to calibrate models of SMBH growth over cosmic time, and many predictions for as yet undetected SMBH populations rest on its inferred slope and amplitude.

\subsubsection{$M_\bullet$ to $M_*$ and $M_{200}$} \label{sec:mbh_mstar_mvir}

We plot the derived relationships between SMBH mass and total stellar mass and SMBH mass and virial mass for the {\sc Romulus} simulations in Figure \ref{fig:mass_relations}.  Here, once again, $M_\bullet$ corresponds to the most massive SMBH associated with each halo.  Only galaxies whose most massive SMBHS are within 2 kpc of galactic centre are shown.  We note that at low-stellar masses, seed masses can comprise the majority of SMBH mass.  Sometimes, multiple seeds merge at high-redshift, since {\sc Romulus} does not impose any minimum distance between seeds.  To compensate for this phenomenon and particulars arising from the seeding methodology adopted, here $M_{\bullet,\mathrm{acc}}$ in these plots corresponds the amount of SMBH mass that is obtained via accretion.  This, therefore excludes all of the seed masses which contribute to the final mass, but includes the accreted portion of every SMBH's mass that may have merged to create the final SMBH.  We confirm that for most SMBHs with mass $>10^7$ M$_{\odot}$ (or a factor of 10 larger than their initial mass), most of the total mass is produced through accretion and their final masses are insensitive to the seed mass assumed.  Points are colour-coded according to each SMBH's Eddington ratio averaged over the past 300 Myr, (a measure of its specific growth rate further discussed in \S\ref{sec:erdfs}).  Those with Eddington ratios above $10^{-2}$ are marked with stars instead of circles.  We overplot relationships between SMBH and stellar mass from \citet{Kormendy&Ho2013} and \citet{Reines&Volonteri2015}.  To derive the correlations with halo mass, we convert these observed relationships via the stellar-to-halo mass relation obtained from abundance matching \citep{Moster+2013}.  Note that the stellar-to-halo mass relation is very uncertain below halo masses of $\sim 10^{11} \ \mathrm{M}_\odot$, and the SMBH-to-stellar mass relation is not observationally measured for stellar masses below $\sim 10^{10} \ \mathrm{M}_\odot$.

Since {\sc Romulus} was calibrated to obtain reasonable SMBH masses for low stellar mass host galaxies, it is encouraging that appropriate masses are obtained for the SMBHs that reside in the most massive haloes, including the $\sim 10^{10} \ \mathrm{M}_\odot$ SMBH in the brighest cluster galaxy (BCG) of {\sc RomulusC}.  There is noticeably more scatter at low-masses than at high-masses, as we would expect from the Central Limit Theorem \citep{Peng2007,Jahnke&Maccio2011}.  Interestingly, there is no indication that the accreted mass departs from a linear relation even below the seed mass of $10^6 \ \mathrm{M}_\odot$, suggesting that galactic inflows on larger scales than the SMBH sphere of influence regulate accretion in low mass galaxies, irrespective of the SMBH mass.

\citet{Reines&Volonteri2015} report separate relationships for elliptical galaxies and broad-line AGN.  By separating {\sc Romulus} galaxies into high- and low- Eddington ratio populations, we find a possible explanation.  Using a multi-wavelength sample of AGN, \citet{Trump+2011} find that a broad line region is only present among AGN with $f_\mathrm{Edd} > 10^{-2}$, which may mark a boundary between radiatively efficient to inefficient accretion disk structures.  As shown in Fig. \ref{fig:mass_relations}, the SMBHs at $z=0$ which fulfill this criterion (marked as stars) have systematically lower masses.  Examining their luminosities, we find that this is not because lower-mass SMBHs have higher accretion rates, but simply because lower-mass SMBHs have higher Eddington ratios for a {\it given} accretion rate.

Turning to halo mass, the $M_\bullet-M_{200}$ relation for {\sc RomulusC} is noticeably offset to the left with respect to {\sc Romulus25}.  However, there is no noticeable difference in the $M_\bullet-M_*$ relationship between the cluster and the field.  This is likely due to tidal stripping, which removes the outer regions of dark matter haloes of cluster members, but is not usually strong enough to impact the total stellar mass.  One might have expected an elevated $M_\bullet-M_*$ relation in {\sc RomulusC} compared to {\sc Romulus25}, since both components assemble earlier in the universe when the ratio $M_\bullet/M_*$ is thought to be higher \citep[e.g.,][]{Yang+2018a}.  Yet as we later show in \S\ref{sec:agnms}, in {\sc Romulus}  the $M_\bullet-M_*$ relation is established earlier in the universe and it does not evolve strongly with redshift.  Furthermore, the main halo of {\sc RomulusC} does not assemble half of its final mass until $z=0.54$, which is relatively recent to expect significant evolution.

In contrast with several other hydrodynamical simulations, there is no indication that SMBH growth is stunted in the lowest-mass galaxies due to supernova feedback.  The EAGLE simulations find that a critical host virial temperature is required for runaway SMBH growth \citep{McAlpine+2018}.  \citet{Bower+2017} suggest that this threshold represents the point beyond which supernova feedback cannot prevent the build-up of gas in the central regions of galaxies. \citet{Dubois+2015} and \citet{Habouzit+2017} similarly find that feedback strong enough to curtail overproduction of stars stunts early SMBH growth in the Seth and SuperChunky simulations respectively.  In the absence of AGN feedback, supernova feedback also regulates SMBH growth in the FIRE simulations \citep{Angles-Alcazar+2017}.  We speculate that the different recipes for accretion and/or feedback used in these simulations may be responsible for these mixed findings.  The relatively high seed mass in {\sc Romulus}, and the propensity for several immediate mergers between seeds, may also play a role.  We further discuss potential explanations of this disagreement later in the \S\ref{sec:discussion}.

\subsubsection{$M_\bullet$ to $\sigma$}

\begin{figure*}
   \centering
   \includegraphics[width=\textwidth]{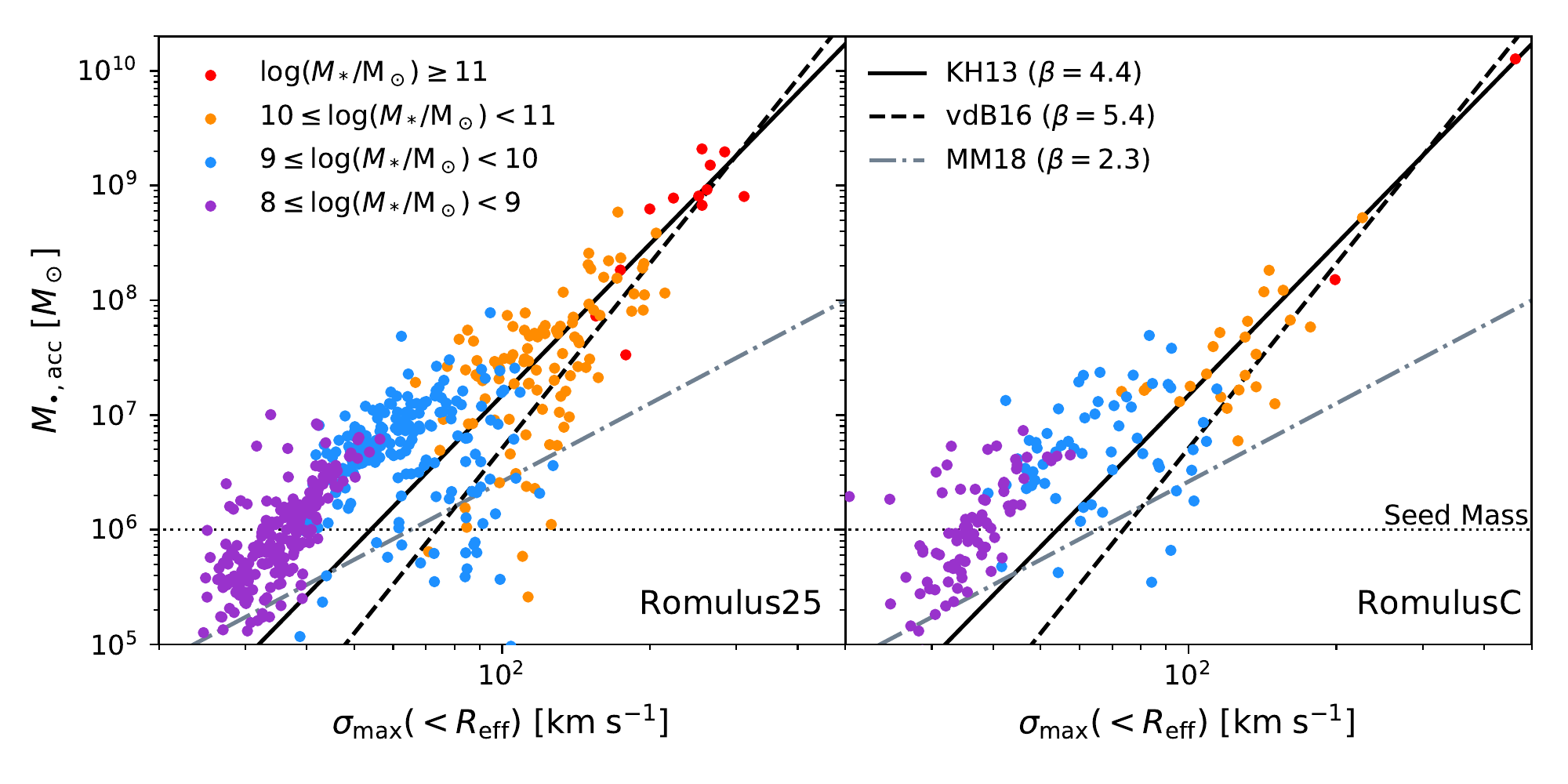}
   \caption{SMBH mass versus host stellar velocity dispersion in {\sc Romulus25}.   Observed relations from \citet{vandenBosch2016}, \citet{Kormendy&Ho2013}, and \citet{Martin-Navarro+2018} are are shown, along with their slopes.  Points are colour-coded according to stellar mass, and more massive galaxies have higher velocity dispersions, as expected.}
   \label{fig:msigma}
\end{figure*}

Some authors have argued that the host velocity dispersion, $\sigma$, is the more fundamental quantity in determining SMBH mass growth \citep{VolonteriPN+2011,Beifiori+2012,vandenBosch2016}.  This quantity reflects not only the host's mass, but also the depth of its potential well.  Indeed, AGN appear to be more common among galaxies which are not only massive but also preferentially compact for their masses \citep{Kocevski+2017,Powell+2017}.  An $M_\bullet-\sigma$ relation can also be theoretically generated by SMBH feedback, with momentum- and energy-driven winds yielding $M_\bullet \propto \sigma^4$ \citep{King2003} and $M_\bullet \propto \sigma^5$ \citep{Haehnelt+1998} relations respectively \citep[see also][]{Natarajan&Treister2009,Zubovas&King2012,King&Pounds2015}. 

We plot the $M_\bullet-\sigma$ relation for {\sc Romulus} galaxies in Figure \ref{fig:msigma} along with the relations observed by \citet{Kormendy&Ho2013}, \citet{vandenBosch2016}, and \citet{Martin-Navarro+2018} for which we have provided the slope of the $M_\bullet-\sigma$ relation, commonly written as $\beta$.  Note that these first two samples are limited to $\sigma \gtrsim 100 \ \mathrm{km} \; \mathrm{s}^{-1}$, while \citet{Martin-Navarro+2018} consider only Seyfert I galaxies.  We estimate $\sigma$ directly from the star particles in the each galaxy.  First, we calculate the effective radius, $R_\mathrm{eff}$, based on the surface brightness profiles of each galaxy.  This is performed in the ``i'' band using cylindrical annuli with the galaxy rotated so that it would be face-on to the observer. The rotation was performed based on the angular momentum of gas, stars and dark matter within the inner 5 kpc of the halo. A single Sersic profile was fit to each galaxy assuming surface brightness cutoff of 32 mag/arcsec$^2$ and a maximum radius of $5\times$ the half-light radius.  Then, we compute $\sigma = \sqrt{ \langle v^2\rangle - \langle v \rangle^2}$.  In order to potentially isolate the velocity dispersion of a bulge that is smaller than the effective radius, we define $\sigma_\mathrm{max}(<R_\mathrm{eff})$ as the maximum velocity dispersion that an observer could infer by enclosing star particles within a maximum radius that varies between 1 kpc (due to our limited resolution) and the effective radius. In practice, we notice little difference between computing $\sigma$ in this manner compared to calculating $\sigma$ for stars enclosed within the effective radius. 

We find that for galaxies with stellar mass above $10^{10} \ \mathrm{M}_\odot$, {\sc Romulus} agrees well with the observed relations for high-mass galaxies.  Even the most massive SMBH in {\sc RomulusC}, that at the centre of the BCG, falls on the relations observed at high masses.  At lower masses, {\sc Romulus} departs from these relations, yielding higher SMBH masses for their hosts' velocity dispersions.  For Seyfert galaxies, \citep{Martin-Navarro+2018} report that the $M_\bullet-\sigma$ relation flattens at these lower host masses, but with a much lower normalization than occurs in {\sc Romulus} \citep{Martin-Navarro+2018}.  In {\sc Romulus}, the $M_\bullet-\sigma$ relation appears to be less fundamental than the $M_\bullet-M_*$ relation.  As we later show in \S\ref{sec:agnms}, this is consistent with the phenomenon that SMBHs grow in tandem with the stellar content of their entire galaxies in these simulations.  Nevertheless, we caution that while the departure from local relations at low-masses is plausible, it is likely quite sensitive to sub-grid recipes for both star formation and SMBH growth, and the resolution of the simulation \citep{Angles-Alcazar+2017}.

\subsection{SMBH Accretion Density}\label{sec:accretion_density}

\begin{figure}
   \centering
   \includegraphics[width=0.45\textwidth]{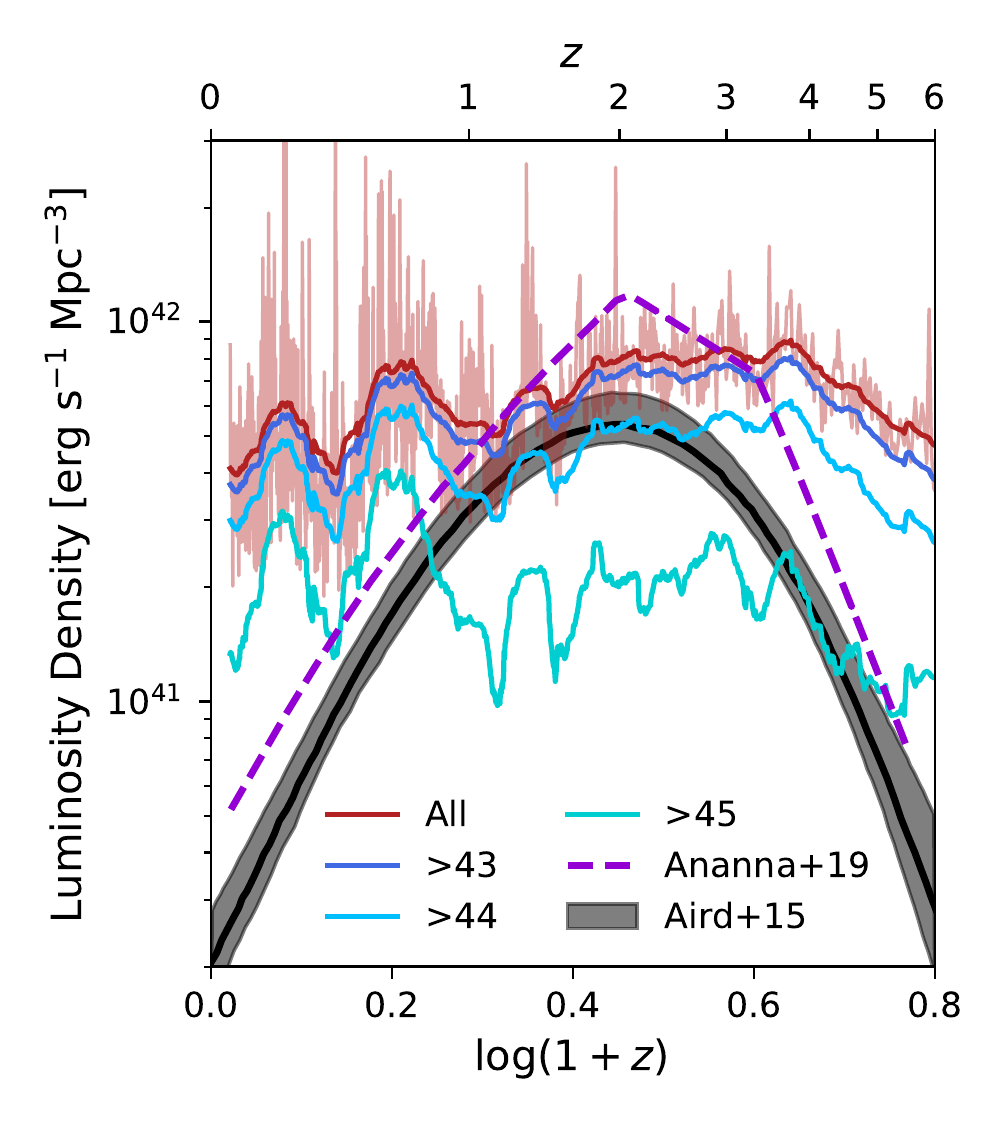}
   \caption{Luminosity density as a function of time in {\sc Romulus25}.  We compare estimates from the population synthesis models of \citet{Aird+2015} and \citet{Ananna+2019}.  Different colours correspond to different bolometric luminosity thresholds as shown in log $\mathrm{erg} \; \mathrm{s}^{-1}$.  Solid lines are smoothed with a boxcar filter, while in light red we display the variability of the original 10 Myr time resolution.  Unlike the population synthesis models, the accretion density is relatively constant for $z<5$ in {\sc Romulus25}.}
   \label{fig:luminosity_density}
\end{figure}

The (25 Mpc)$^3$ volume of {\sc Romulus25} does not allow for the robust predictions of AGN luminosity functions, since the knee of the luminosity function falls below the minimum number density that can be probed in this box-size.  In order to compare the assembly history of the {\sc Romulus25} SMBHs with the redshift-evolving AGN census, we instead calculate the bolometric luminosity density.  This is the integral of the luminosity function and captures the total amount of accretion that occurs at a given epoch.  This quantity is unaffected by any time-variability that is not resolved by the simulation.  Variability may alter the shape of luminosity functions, but not its integral, making it a robust quantity to compare with observational data. 

We plot the AGN bolometric luminosity density in {\sc Romulus25} in Fig. \ref{fig:luminosity_density} for four different bolometric luminosity thresholds corresponding to $0$, $10^{43}$, $10^{44}$, and $10^{45} \ \mathrm{erg} \; \mathrm{s}^{-1}$.  For readability, the solid curves have been smoothed using a boxcar filter with a full width of $\Delta \log (1+z) = 0.05$.  In light red, we also plot the original $\Delta t = 10 \ \mathrm{Myr}$ resolution with no luminosity threshold.  There is substantial stochasticity due to the relatively small box-size of {\sc Romulus25}.  Along the same lines, the $10^{45} \mathrm{erg} \; \mathrm{s}^{-1}$ curve also reveals that a few rare, luminous objects drive the shape of the volume-averaged variability.

We then compare to the bolometric luminosity density estimated from two different population synthesis models based on X-ray observations \citep{Aird+2015,Ananna+2019}. These measurements are sensitive to absorbed AGN and include an estimate of the Compton-thick contribution required in order to fit the cosmic X-ray background spectrum.  The model of \citet{Ananna+2019} sits vertically offset from that of \citet{Aird+2015} due to the higher inferred fraction of Compton thick sources.  There is an interesting disagreement between these population synthesis models and {\sc Romulus25} in terms of both the shape and normalization of these curves.  In {\sc Romulus25}, the luminosity density matches the observations only during the epoch of peak AGN activity and stays remarkably constant throughout cosmic time.  We offer two possible explanations.  In \citet{Tremmel+2017}, it was shown that {\sc Romulus25} similarly overestimates the density of star-formation at $z \lesssim 1$ and possibly $z \gtrsim 3$.  The quenched fraction in {\sc Romulus} is systematically low, particularly at high-masses \citep{Tremmel+2019}.  As we shall show in \S\ref{sec:agnms}, AGN activity follows the star formation rate in {\sc Romulus}, and this is likely the cause of the overestimate of the AGN luminosity density as well.  At high-redshift, it is also possible that current surveys miss fainter and/or more obscured populations.  As we show, the total luminosity density in {\sc Romulus25} is sensitive to the threshold above which we include SMBH activity.

\subsection{The AGN Main Sequence}\label{sec:agnms}

We now explore the relationships between SMBHs and their hosts to determine how SMBHs and their galaxies co-evolve.  By stacking X-ray observations of star-forming galaxies, \citet{Mullaney+2012} reported a linear relation between the BHAR and the SFR, the so-called hidden ``AGN Main Sequence.''  Subsequent studies stacking star-formation rates of X-ray AGN host galaxies appeared to refute this picture, except for perhaps the most luminous AGN \citep{Rosario+2012,Stanley+2015,Azadi+2015,Ramasawmy+2019}.  It is now thought that an AGN varies on shorter timescales than its host's star formation \citep{Hickox+2014}, washing out correlations if one stacks on AGN luminosity instead of star formation rate \citep{Azadi+2015,Lanzuisi+2017}.  In addition, the relation obtained depends on the shape of the bivariate distribution, how objects are selected, and how they are binned \citep{Volonteri+2015b,McAlpine+2017}.  Recently, some authors have claimed that stellar mass, rather than SFR, is the more fundamental quantity governing the BHAR \citep{Yang+2017,Fornasini+2018}, except perhaps among bulge-dominated galaxies \citep{Yang+2019}.  \citet{Aird+2019} found relationships between {\it specific} accretion rates and the SFR, which was elevated for quiescent galaxies.

\begin{figure*}
   \centering
   \includegraphics[width=\textwidth]{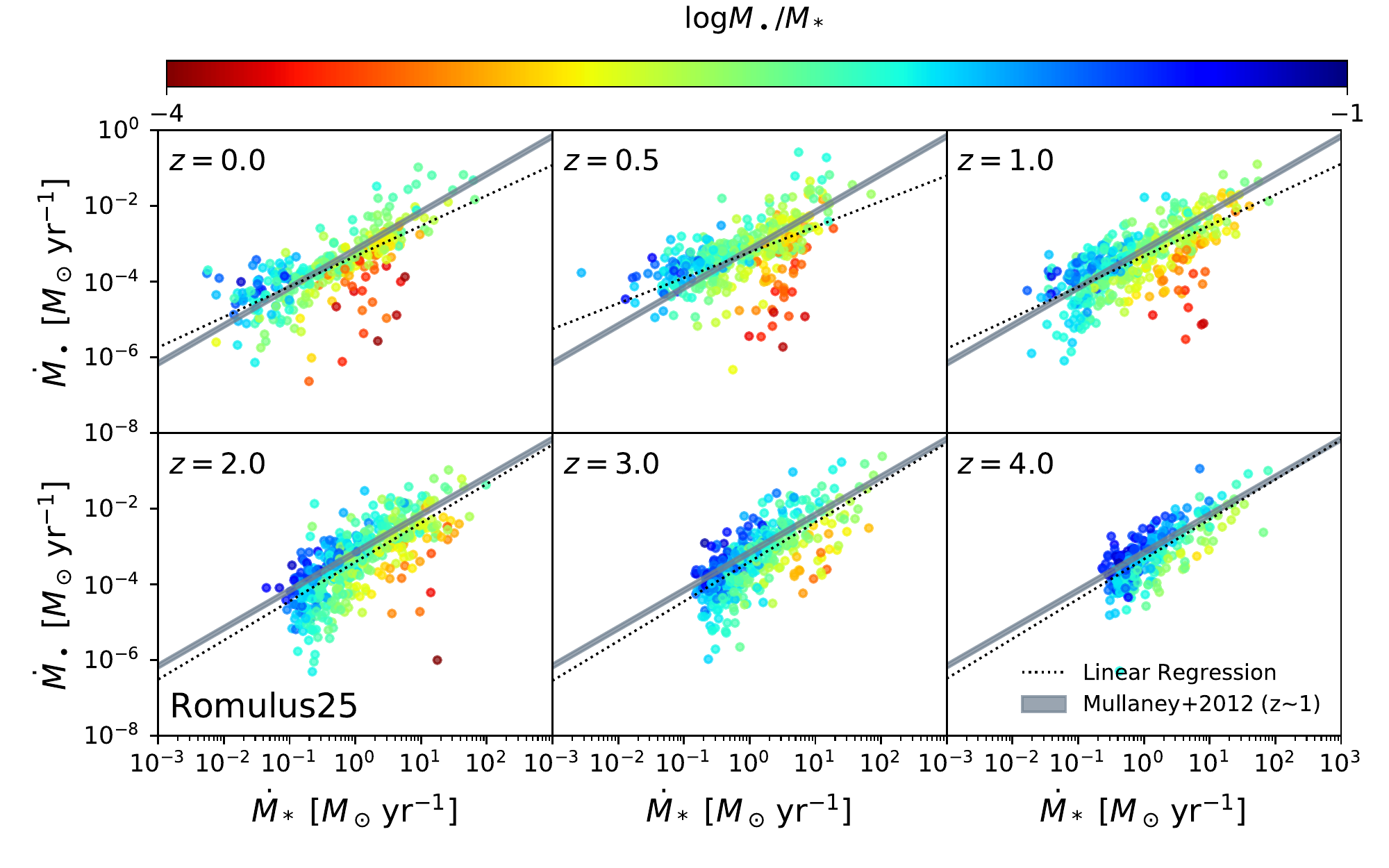}
   \caption{BHAR versus SFR for star-forming galaxies in {\sc Romulus25}. Both quantities are averaged over the past 300 Myr.  In grey, we overplot the relation observed at $z=1$ from stacked star-forming galaxies \citep{Mullaney+2012}.  This relation agrees well with a linear regression, shown as a black dotted line, regardless of redshift.  Points are colour-coded according to the SMBH to stellar mass ratio.}
   \label{fig:agnms_R25}
\end{figure*}

We plot the relationship between BHAR and SFR {\it for the star-forming galaxies} of {\sc Romulus25} in Figure \ref{fig:agnms_R25}.  We include only galaxies with stellar masses of at least $10^8 \ \rm{M}_\odot$ that are no more than 1 dex below the star-forming sequence.  The BHAR of the most massive SMBH is shown, and we exclude cases where the SMBH is greater than 2 kpc from the halo centre.  Finally, in order to mitigate AGN variability and compare both quantities on the same timescale, we average both the BHAR and the SFR over the past 300 Myr.  We plot a simple linear regression in logarithmic space (a power law fit) as a black dotted line.  In grey, we overplot the relation observed at $z=1$ by \citet{Mullaney+2012}.  {\sc Romulus25} agrees remarkably well with this relation, even at much higher and lower redshifts.  Points are colour-coded according to the ratio between SMBH mass and stellar mass.  This reveals a clear vertical gradient, indicating that some of the scatter is correlated with SMBH to stellar mass ratio.  Note that in the least-massive galaxies, mass ratios are artificially large due to the large seed mass.  We will return to this correlation in \S\ref{sec:statistical_search}.

\begin{figure}
   \centering
   \includegraphics[width=0.5\textwidth]{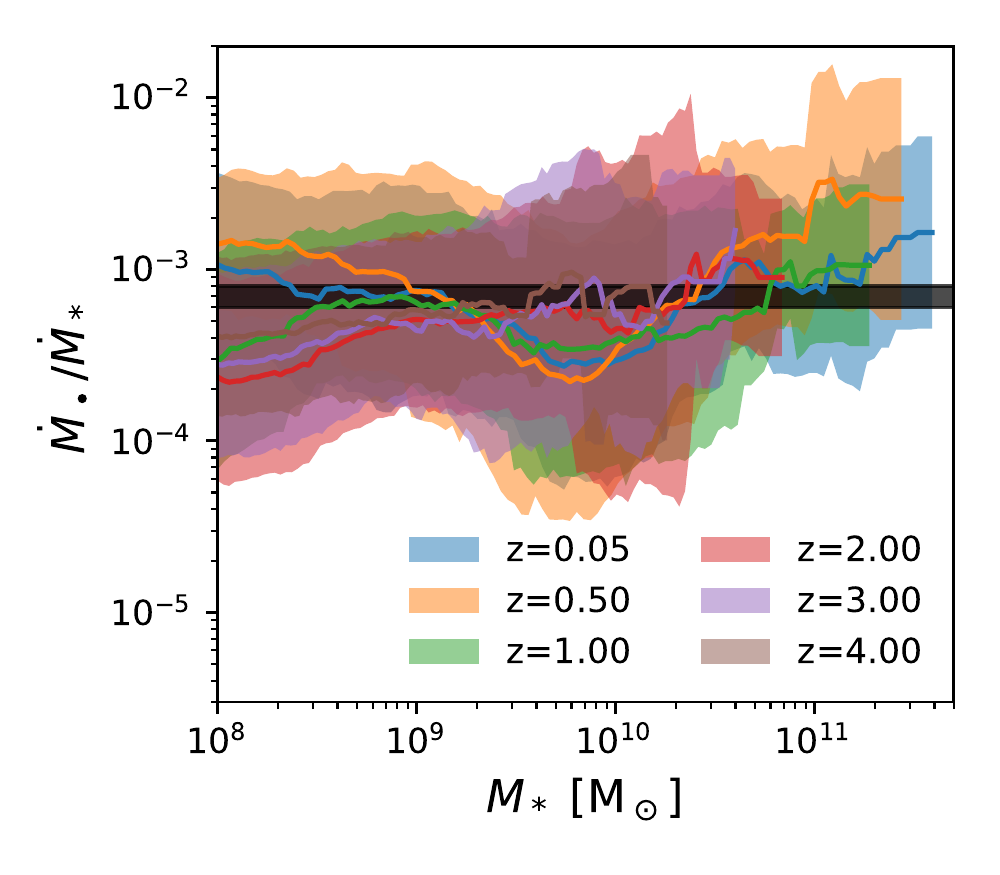}
   \caption{The ratio of BHAR to SFR in {\sc Romulus25} as a function of stellar mass.  There is no trend as a function of either stellar mass or redshift.  The black region corresponds to the $z\sim 1$ relation observed by \citet{Mullaney+2012}.}
   \label{fig:agnms_mass_dependence}
\end{figure}

The relationship between BHAR and SFR does not appear to change as a function of redshift or stellar mass.  This is better illustrated by Fig. \ref{fig:agnms_mass_dependence}.  Here, we represent the same data in Fig. \ref{fig:agnms_R25}, and compute the moving average $\dot{M}_\bullet/\dot{M}_*$ and its standard deviation using a boxcar filter with total width of $\Delta \log (M_*/\mathrm{M}_\odot) = 0.6$.  The black region represents the \citet{Mullaney+2012} $z\sim 1$ relation, a constant ratio.  {\sc Romulus25} is consistent with this relation at all redshifts and stellar masses.  At the highest masses, there might be a slight upturn, especially at lower redshifts, albeit within the scatter of these relations.  This may be related to the fact that there do not exist galaxies with low values of $M_\bullet/M_*$ for stellar masses above $10^{11} \ \mathrm{M}_\odot$ (see Fig. \ref{fig:mass_relations}).

\begin{figure*}
   \centering
   \includegraphics[width=\textwidth]{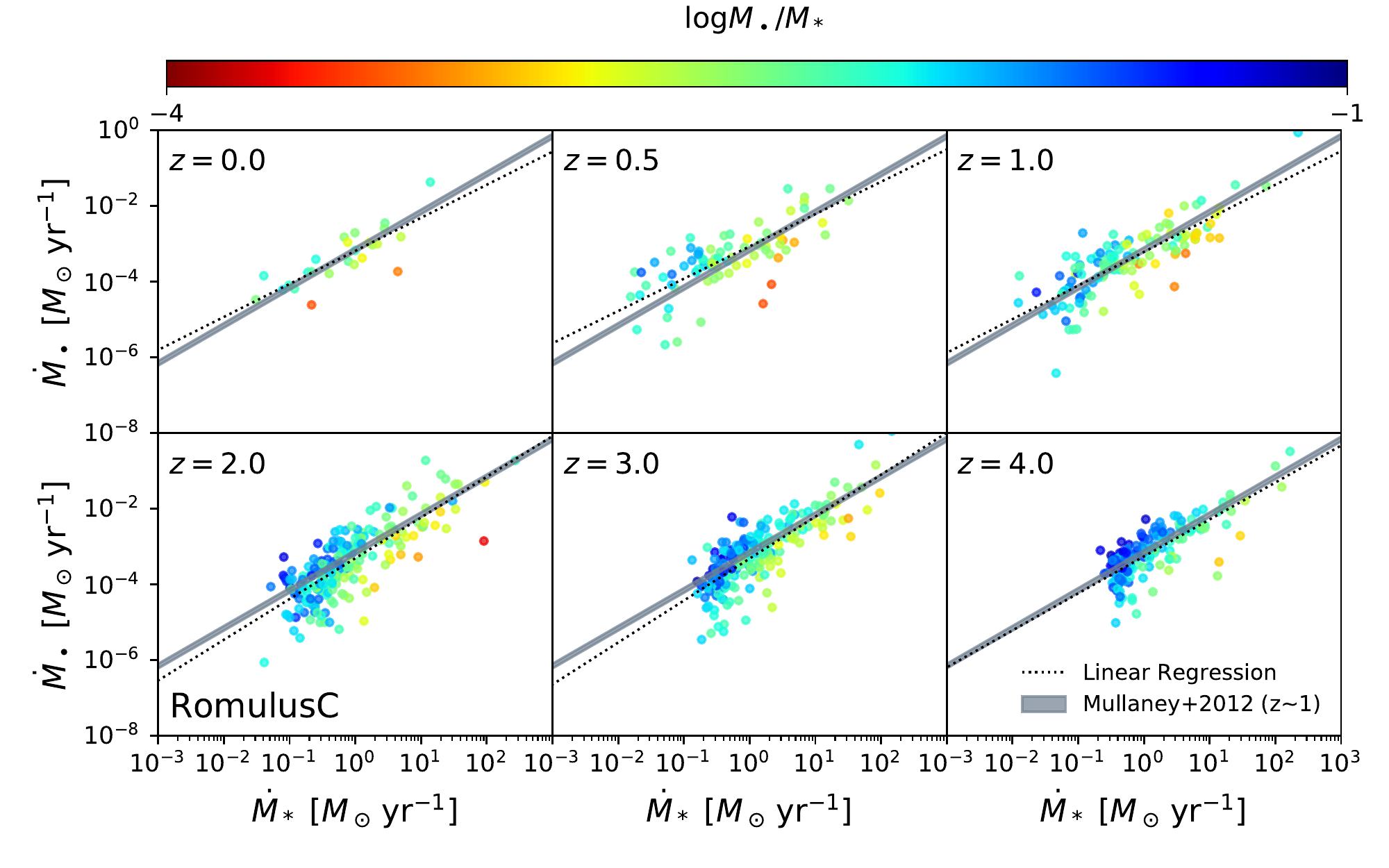}
   \caption{As in Figure \ref{fig:agnms_R25}, but for {\sc RomulusC}.  There are fewer points at low-redshift because only star-forming galaxies are included.  Since the same relation from \citet{Mullaney+2012} can describe both these cluster galaxies and the field galaxies in Fig. \ref{fig:agnms_R25}, there appears to be no difference between the field and the cluster environments.}
   \label{fig:agnms_RC}
\end{figure*}

In Fig. \ref{fig:agnms_RC}, we repeat this analysis for {\sc RomulusC} to search for differences between the field and cluster.  Note that there are many fewer galaxies included at low-redshift because the majority of cluster-members become quenched.  Remarkably, despite the very different intergalactic environments, the galaxies which remain on the star forming sequence are still consistent with the \citet{Mullaney+2012} relation.  Hence, there appears to be no difference between the cluster and the field.  Previous work on {\sc RomulusC} has determined that although AGN are suppressed overall in the cluster, there still exists a high-Eddington ratio population at low-redshift \citep[see Figure 18 in][]{Tremmel+2019}.  Our interpretation is that quenching is demonstrably more efficient in the cluster environment, but before it is quenched, a galaxy and its SMBH assemble independently of the larger intergalactic environment.

\begin{figure*}
   \centering
   \includegraphics[width=\textwidth]{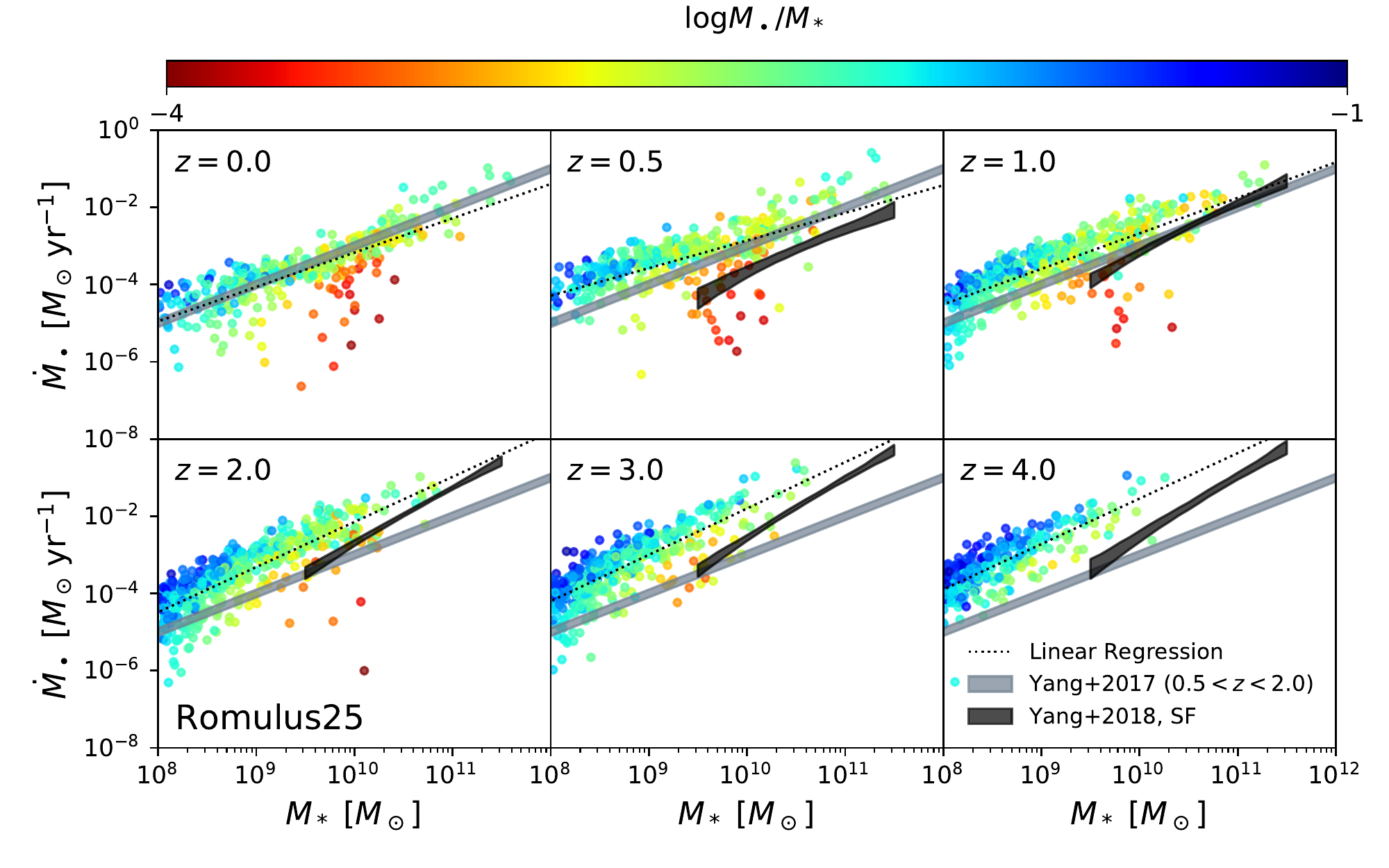}
   \caption{As in Figure \ref{fig:agnms_R25}, but for stellar mass instead of SFR.  Although the relationship is tighter than with SFR, a redshift dependence is required.  A relationship derived from 18,000 galaxies in the CANDELS/GOODS-South field is overplotted in grey \citep{Yang+2017}, while a redshift-evolving relationship based on CANDELS and COSMOS galaxies is shown in black \citep{Yang+2018a}.  Note that {\sc Romulus} tends to over-predict the BHAR compared to observational data, as we previously reported in Fig. \ref{fig:luminosity_density}.}
   \label{fig:agnms_Mstar}
\end{figure*}

Finally, we also test the hypothesis that the BHAR depends on the stellar mass more strongly than the SFR \citep{Yang+2017,Yang+2018a}.  In Figure \ref{fig:agnms_Mstar}, we repeat our analysis of the AGN Main Sequence in {\sc Romulus25}, replacing the independent variable with stellar mass.  It is visually apparent that the scatter of the relation is smaller with stellar mass than with SFR, {\it but this is true only at fixed redshift}.  In {\sc Romulus}, the normalization of the BHAR-$M_*$ correlation increases with increasing redshift, but there is no evidence of redshift evolution in the relationship between BHAR and SFR (see Fig. \ref{fig:agnms_mass_dependence}).  We therefore arrive at the nuanced conclusion that stellar mass is a better predictor of the BHAR at fixed redshift, but SFR is a better predictor of the BHAR across a large range of redshifts.  This motivates a more statistically rigorous multi-variable analysis, which we undertake in the following section.

\subsection{A Statistical Search for the Most Fundamental Relation}\label{sec:statistical_search}

To search for the most fundamental relations between the BHAR and global galaxy properties, we preform multi-linear regressions in logarithmic space for a variety of models.  Our metric of choice to discriminate between models is the Bayesian Information Criterion (BIC), which penalizes models for adding unnecessary parameters \citep{Schwarz1978}.  Making the assumption that scatter about these relations is Gaussian, we express the BIC as 
\begin{equation}
    \mathrm{BIC} = n \ln{\widehat{\sigma^2_e}} + k \ln(n)
\end{equation}
\noindent where $n$ is the number of data points, $\widehat{\sigma^2_e}$ is the error variance, and $k$ is the number of free parameters.  The raw value of the BIC does not carry significant meaning, but a change in its value does.  Preferred models have smaller (in our case, more negative) values of the BIC.

Our multi-linear models take the form
\begin{equation}
\log_{10} (\dot{M}_\bullet) = \log_{10} (\dot{M}_{\bullet,0}) + \sum C_x \log_{10}(x)
\label{eqn:fit_form}
\end{equation}
\noindent where $x$ represents any independent variable, such as $(1+z)$ or $\dot{M}_*$. BHAR and SFR both have units of $\mathrm{M}_\odot \; \mathrm{yr}^{-1}$.  We fit the values obtained in \S\ref{sec:agnms} for snapshots $z \in \{0.05,0.5,1.0,2.0,3.0,4.0\}$, epochs which are sufficiently separated in time to be treated as independent points despite our 300 Myr smoothing.  In addition to the processing described in the previous section, we also restrict our fits to galaxies with stellar masses of at least $10^9 \ \rm{M}_\odot$.  This is done because the seed mass leads to artificially high values of the SMBH-to-stellar mass ratio in low-mass galaxies.

We test several different models, whose independent variables are listed here:
\begin{itemize}
    \item Model A:  $M_*$
    \item Model B:  $\dot{M}_*$
    \item Model C:  $M_*$, $(1+z)$
    \item Model D:  $\dot{M}_*$, $q$
    \item Model E:  $\dot{M}_*$, $q$, $f_\mathrm{gas}$
    \item Model F:  $\dot{M}_*$, $q$, $f_\mathrm{cold}$
\end{itemize}
\noindent where here we define $q = M_\bullet/M_*$, $f_\mathrm{gas} = M_\mathrm{gas} / (M_\mathrm{gas} + M_*)$, and $f_\mathrm{cold} = M_\mathrm{gas,cold} / M_\mathrm{gas}$.  Gas is designated as ``cold'' if it is below $2\times10^4$ K.  Best-fit values are provided in Tab. \ref{tab:AGNMS}, along with several statistics of fit.  As we show, these models are sorted in order of increasing preference.

\subsubsection{Stellar Mass, Star-formation Rate, and Other Parameters}

\begin{table*}
\begin{tabular}{lcccccccccc}
\hline
        & $\log_{10}(\dot{M}_\bullet,0)$ & $C_{\dot{M}_*}$ & $C_{M_*}$ & $C_{q}$ & $C_{(1+z)}$ & $C_{f_\mathrm{gas}}$ & $C_{f_\mathrm{cold}}$ & $\sqrt{\widehat{\sigma^2_e}}$ & $R^2$ & BIC     \\ \hline
Model A & -9.78                          &                 & 0.709     &         &             &                      &                       & 0.741                         & 0.193 & -606.5  \\
Model B & -3.20                          & 0.763           &           &         &             &                      &                       & 0.687                         & 0.308 & -765.3  \\
Model C & -12.8                          &                 & 0.948     &         & 2.15        &                      &                       & 0.606                         & 0.462 & -1018.4 \\
Model D & -0.0712                        & 0.946           &           & 1.21    &             &                      &                       & 0.490                         & 0.648 & -1458.7 \\
Model E & 0.0418                         & 0.977           &           & 1.29    &             & -0.630               &                       & 0.482                         & 0.660 & -1489.4 \\
Model F & -0.386                         & 0.904           &           & 1.17    &             &                      & -0.434                & 0.464                         & 0.684 & -1564.9 \\ \hline
\end{tabular}
\caption{Best-fit values for multi-linear regressions to the relationship between $\dot{M}_\bullet$ and global galaxy properties, defined in equation \ref{eqn:fit_form}.  In these simulations, $\dot{M}_*$ is a better predictor of $\dot{M}_\bullet$ than $M_*$. Adding $q$ as a second parameter greatly improves the model, while adding gas-related third parameters mildly improve the model as well. \label{tab:AGNMS}}
\end{table*}

Models A and B attempt to relate the BHAR to solely the stellar mass and the star-formation rate respectively, with no redshift evolution.  Referring to Table \ref{tab:AGNMS}, these models have $R^2$ values of 0.19 and 0.31 respectively.  As we have seen in Fig. \ref{fig:agnms_Mstar}, the poor $R^2$ value in Model A is due to the redshift-evolution.  Model C includes a $(1+z)$ dependence to this $M_*$-dependent model, which significantly increases its $R^2$ value to 0.46.  However, a competing model with the same number of parameters, Model D, yields a much higher $R^2$ value of 0.65 with two physical parameters:  $\dot{M}_*$ and $q$.  We conclude that $\dot{M}_*$ is a more fundamental parameter than $M_*$, and that $q$ provides a significant improvement on top of that single-parameter.  To be more succinct, we find that the models of increasing quality are:  $M_* < \dot{M}_* < M_*, (1+z) < \dot{M}_*, q$.  The BIC decreases by at least 100 for each of these steps, while a difference of only 10 is considered strong evidence in favor of the model with the lower BIC.

Although $q$ has been shown to be a statistically significant additional parameter, we note that this relationship is not necessarily causal.  There are physical reasons to expect a near-linear relationship with $q$.  Recall that under the assumption of modified Bondi accretion, the accretion rate should increase with SMBH mass, as we observe.  In addition, higher values of SMBH mass permit higher values of the BHAR due to the Eddington limit, although we show in \S\ref{sec:erdfs} that the Eddington limit is probably not a major limitation in these simulations.  On the other hand, the relationship with $q$ may simply reflect the fact that galaxies which have successfully fuelled their SMBHs in the past are more likely to continue doing so.  In this latter interpretation, the correlation with $q$ is a reflection of more fundamental, unknown galactic conditions that remain stable over timescales of at least hundreds of millions of years.

Having established a model based on $\dot{M}_*$ and $q$ (Model D), we search for a third parameter.  Two more quantities which one might expect to correlate with $\dot{M}_\bullet$ are the baryonic gas fraction, $f_\mathrm{gas}$, and the fraction of gas which is cold, $f_\mathrm{cold}$, which we explore in models E and F.  The addition of either of these parameters does significantly improve upon Model D on the basis of the BIC, with $f_\mathrm{cold}$ preferred over $f_\mathrm{gas}$.  Interestingly, we find that the coefficients for either of these parameters is {\it negative}.  That is, for a given SFR and mass ratio, larger values of either the gas fraction or the cold gas fraction correspond to {\it less} black hole accretion.  These anti-correlations persist even if all satellite galaxies are removed, defined in this case to be galaxies within the virial radius of another galaxy with higher stellar mass.  This disfavours environmental processes as the drivers of these anti-correlations.  Instead, this may reflect SMBH feedback:  SMBHs which are accreting more efficiently can either heat the gas or evacuate it from their host galaxies.

\subsubsection{A Closer Look at the Scatter}

\begin{table}
\begin{tabular}{lc}
\hline
\textbf{Method}                         & \textbf{$\sqrt{\widehat{\sigma^2_e}}$} \\
\hline
Smooth over 30 Myr.                    & 0.759                 \\
Smooth over 300 Myr.                   & 0.628                 \\
As above, and remove quenched galaxies. (Final.) & 0.490       \\         
\hline
\end{tabular}
\caption{Scatter obtained for Model D ($\dot{M}_*$ and $q$) with different methodologies.  Averaging over longer time periods and removing quenched galaxies substantially tightens the relationship. \label{tab:AGNMS_scatter}}
\end{table}

In Tab. \ref{tab:AGNMS_scatter}, we report how the scatter in Model D ($\dot{M}_*$ and $q$) decreases by both averaging over a longer timescale and removing quenched galaxies.  Both of these steps are important in obtaining a tighter relation.  The first of these steps helps to mitigate AGN variability.  As we have illustrated in Fig. \ref{fig:luminosity_density}, there is significant stochasticity on the 10 Myr timescales over which BHAR data are saved.  The second step, removing quenched galaxies, tends to remove outliers.  We comment that these quenched galaxies preferentially have high BHARs for their SFRs.

\subsection{Cross-correlation Analysis}\label{sec:cross-correlation}

In previous sections, we have established a preferred ratio between the BHAR and SFR.  Here, we search for a cross-correlation between the BHAR and SFR with a variable time-lag.  This would help determine if the {\it shapes} of these curves are interdependent.  To this end, we calculate the Pearson correlation coefficient ``$r$,'' defined for a sample of paired data $\{(x_i,y_i)\}$ as

\begin{equation}
    r_{x,y} = \frac{\sum_i (x_i - \bar{x})(y_i - \bar{x})}{\sqrt{\sum_i (x_i-\bar{x})^2} \sqrt{\sum_i(y_i-\bar{y})^2}}
\end{equation}

\noindent where $\bar{x}$ and $\bar{y}$ denote the mean of $\{x\}$ and $\{y\}$.  Perfect correlation corresponds to $r=1$, while perfect anti-correlation corresponds to $r=-1$.  We emphasize that this calculation removes the means of each sample, isolating for each galaxy how the {\it shape} of the BHAR curves influence the SFR curves, or vice-versa.

Starting at a given redshift, we first compute the BHAR and SFR for each galaxy above a stellar mass of $10^8 \ M_\odot$ going back 1 Gyr.  To mimic the selections employed in \S\ref{sec:agnms}, we then remove all galaxies which fall 1 dex below the star forming sequence.  Next, we calculate this value between BHAR and SFR while adding a variable time-lag between the two quantities.  To avoid adding spurious features, we ensure that we do not assume that these time series are periodic.

\begin{figure*}
   \centering
   \includegraphics[width=\textwidth]{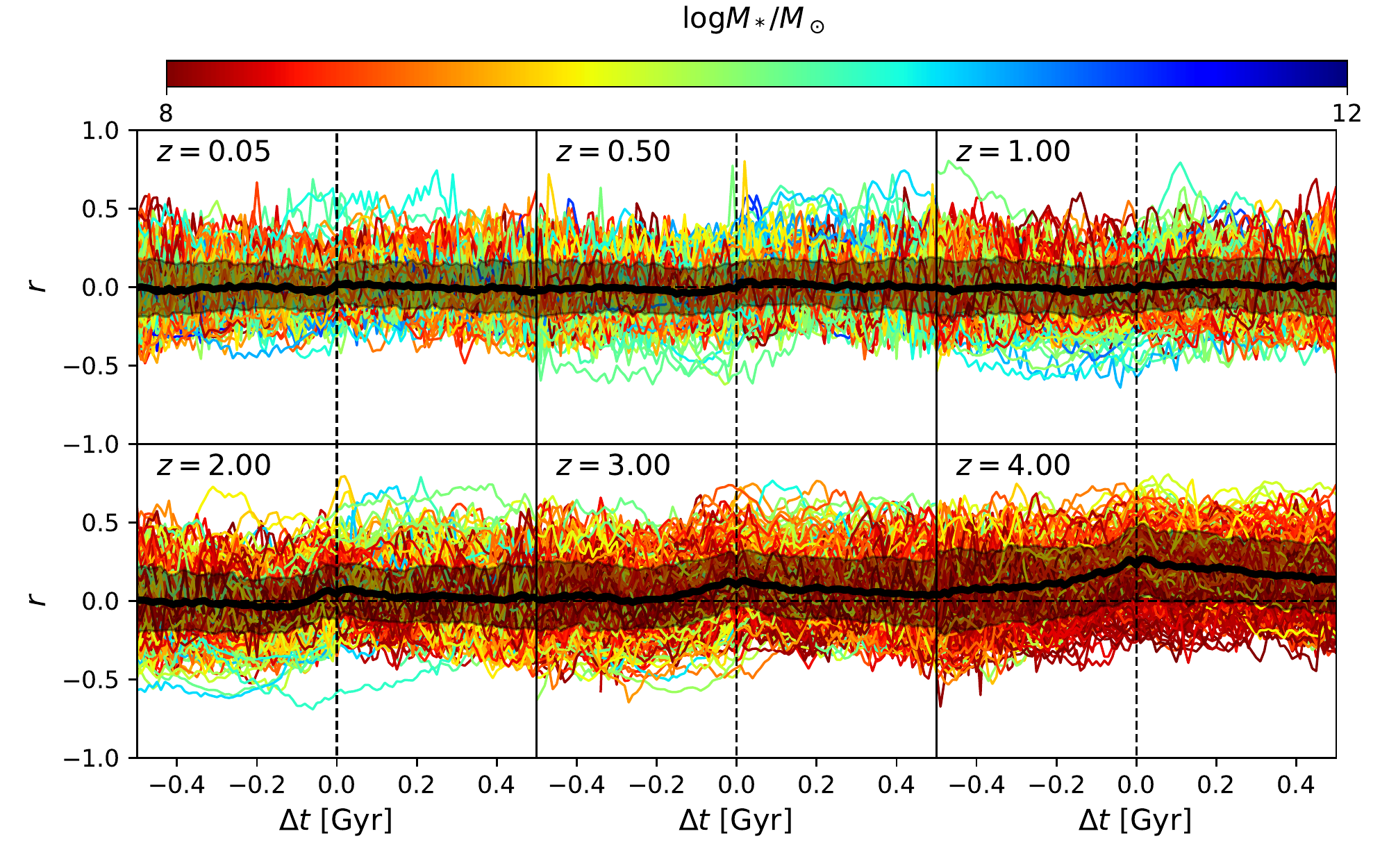}
   \caption{Cross-correlations between BHAR and SFR.  The black solid line corresponds to the median among all galaxies, the black filled region encloses the 1$\sigma$ region, and each overlaid curve corresponds to an individual galaxy that is colour-coded according to its stellar mass.  There is no cross-correlation between the two quantities either instantaneously or with a time-lag.  Although there is a preferred ratio between BHAR and SFR, variations in one do not predict variations in the other.  Note that the mild preference for positive correlations at high-redshift on all time-scales is likely due to the tendency for both BHAR and SFR to significantly increase over 1 Gyr at this epoch.}
   \label{fig:cross_correlation}
\end{figure*}

In Fig. \ref{fig:cross_correlation}, we plot the results for six different starting redshifts.  Cross-correlation functions for each galaxy are overlaid in the background with colours encoding the stellar mass.  The x-axis corresponds to the value of the variable time lag, with positive values representing the BHAR leading the SFR.  In black, we plot the 1$\sigma$ region among all galaxies, while the solid black curve represents the median.

On average, there is no cross-correlation between the BHAR and SFR with time lags up to 0.5 Gyr.  This means that although there is a preferred ratio between the BHAR and SFR as established in sections \ref{sec:agnms} and \ref{sec:statistical_search}, a fluctuation in one value does not predict a similar fluctuation in the other.  At high-redshift, it appears as if positive cross-correlations are preferred at all time-scales.  However, we believe this is likely due to the fact the universe changes significantly over 1 Gyr at these redshifts, increasing in both BHAR and SFR, which naturally leads to a positive cross-correlation.  This negative result does not change if we first take the logarithm of both quantities, if we restrict to narrower selections of stellar mass, or if we first smooth both quantities to a common variability timescale of 30 Myr.  These results imply that while there is a preferred ratio between the BHAR and SFR, each quanity varies independently.

\subsection{Eddington Ratio Distributions}\label{sec:erdfs}

Like most cosmological simulations, the {\sc Romulus} simulations cap the SMBH accretion rate at the Eddington limit.  The Eddington limit is the maximum luminosity at which an accreting object can shine, assuming local force-balance between gravity and radiation pressure \citep{Eddington1926}.  This is given by

\begin{equation}
    L_\mathrm{Edd}(M_\bullet) = \frac{4\upi GM_\bullet m_p c}{\sigma_T} \approx 1.3 \times 10^{44} \bigg( \frac{M_\bullet}{10^6 \ \rm{M}_\odot}\bigg) \mathrm{erg} \; \mathrm{s}^{-1}
\end{equation}

\noindent where $m_p$ is the proton mass and $\sigma_T$ is the Thompson cross-section.  A SMBH's Eddington ratio is defined $f_\mathrm{Edd} = L_{\bullet,\mathrm{bol}}/L_\mathrm{Edd}$.  

The extent to which the Eddington limit influences SMBH assembly remains an important topic, with profound implications for the growth of the highest-redshift quasars.  The earliest quasars at $z\sim 6-7$ have broad-line masses that require uninterrupted Eddington-limited accretion since the seeding epoch for stellar mass seeds, or nearly uninterrupted Eddington-limited accretion for heavy seeds \citep{Haiman&Loeb2001,Volonteri+2015a,Pezzulli+2016}.  The allowance of super-Eddington accretion helps assemble these masses with shorter duty cycles of feeding events.  State-of-the-art general relativistic radiative magneto-hydrodynamical (GRRMHD) simulations have shown that super-Eddington mass accretion rates are sustainable at the accretion disk scale \citep{Jiang+2014,McKinney+2015,Dai+2018}.  Yet despite this, the observed population of AGN does appear to strictly obey the Eddington limit \citep{Wu+2015}.  Perhaps in reality, galactic conditions on larger scales than those accessible to GRRMHD simulations do not permit super-Eddington inflows.

\begin{figure*}
   \centering
   \includegraphics[width=\textwidth]{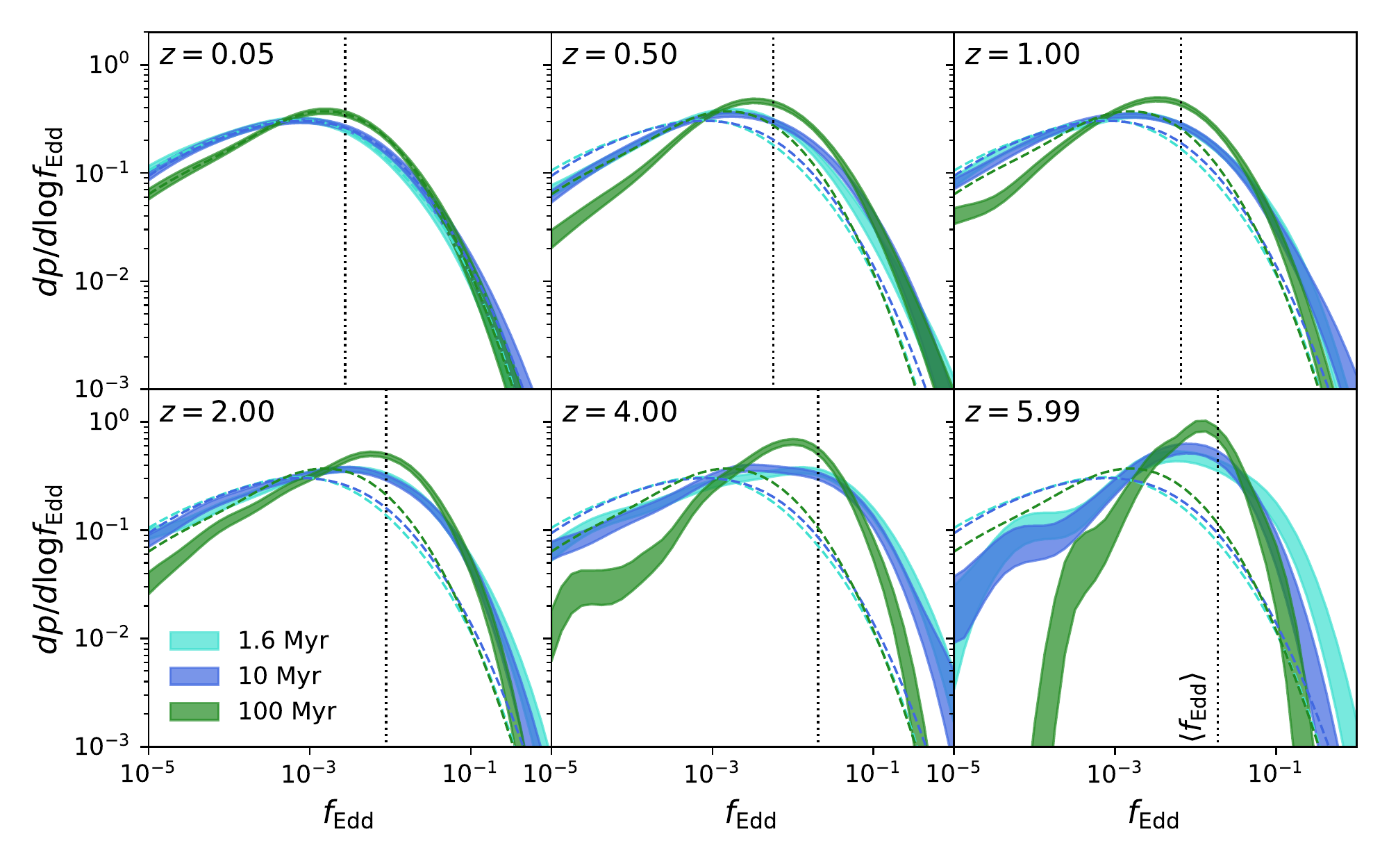}
   \caption{Eddington ratio distributions as a function of redshift and time-interval.  The dotted line in each panel shows the $z=0$ distributions.  Eddington ratios rise as redshift increases.  The timescale over which the Eddington ratio is computed also matters, with shorter timescales producing more frequent extreme low- and high-values.}
   \label{fig:erdf}
\end{figure*}

We investigate, therefore, the distribution of Eddington ratios in {\sc Romulus25}, plotted in Figure \ref{fig:erdf}.  We include the most massive SMBH in each halo above $10^8 \ \mathrm{M}_\odot$, even if its accretion rate is zero.  We plot Eddington ratio distributions where the BHAR is averaged over three different timescales, 1.6 Myr, 10 Myr, and 100 Myr.  Our distributions are derived using a Gaussian kernel density estimation technique.  (Note that the shape of the high-Eddington ratio falloff is due entirely to assuming a Gaussian kernel.)  The dashed coloured lines in each panel correspond to the $z=0$ values to guide the eye, while the black dotted line demarcates the mean log Eddington ratio, $\langle f_\mathrm{Edd} \rangle$, on 1.6 Myr timescales.

As redshift increases, the Eddington ratio distribution shifts towards higher values.  This is as expected, since specific star formation rates also increase with redshift, likely reflecting the fact that these galaxies are relatively more gas-rich overall.  The mean Eddington ratio rises from $10^{-2.6}$ at $z=0.05$ to $10^{-1.7}$ at $z=6$.  Overall, this behaviour is consistent with a trend $\langle f_\mathrm{Edd} \rangle \propto (1+z)$.  We find that the Eddington ratio distributions we derive depend somewhat on the timescale over which they are averaged.  The shorter the timescale, the more frequent $f_\mathrm{Edd} > 0.1$ events are.  (See the $z=4,6$ panels.)  In other words, {\sc Romulus25} does not sustain Eddington-limited accretion flows for 100 Myr periods.  AGN Variability between 1 and 100 Myr produces more AGN at both higher and lower accretion rates than the mean of $10^{-2.5}$.  Unresolved variability on timescales shorter than 1.6 Myr, including ``flicker,'' may continue to modify these distributions when comparing simulation values to instantaneous values that can be observed.

Even at $z=6$, these distributions begin to fall off well before the Eddington limit.  Only $2.5 \pm 1.0$ per cent of SMBHs are accreting at $f_\mathrm{Edd} > 0.1$ on 1.6 Myr timescales at $z=6$.  This suggests that {\it for the typical galaxy}, the Eddington limit is not a relevant barrier to SMBH growth.  Rather, intragalactic astrophysics operating on larger scales sets the accretion rate.  We emphasize, however, that this may not be the case for quasars selected in flux-limited samples.

\subsection{Mergers Do Not Affect the Accretion Rate}\label{sec:triggers}

High-resolution, idealized merger simulations indicate that galaxy mergers can trigger AGN by disturbing the gravitational potential, causing gas to shock against itself and lose angular momentum \citep{DiMatteo+2005,Capelo&Dotti2017}.  Cosmological simulations have mostly found little to no connection with mergers \citep{McAlpine+2018,Steinborn+2018,Martin+2018}.  That said, high-resolution studies indicate that much of the angular momentum loss occurs within 50 pc of the SMBH \citep{Capelo+2015}, which is currently inaccessible to cosmological simulations.  Morphological studies of the hosts of X-ray or optically-selected AGN have also struggled to find a significant connection between merger and AGN \citep{Cisternas+2011,Mechtley+2016,Villforth+2017}.  Infra-red selected AGN, however, tend to be found more often in merging systems \citep{Glikman+2015,Kocevski+2015,Fan+2016,Ricci+2017,Donley+2018,Trakhtenbrot+2018}, suggesting that late-stage merger-driven SMBH growth could be heavily obscured.

\begin{figure*}
   \centering
   \includegraphics[width=\textwidth]{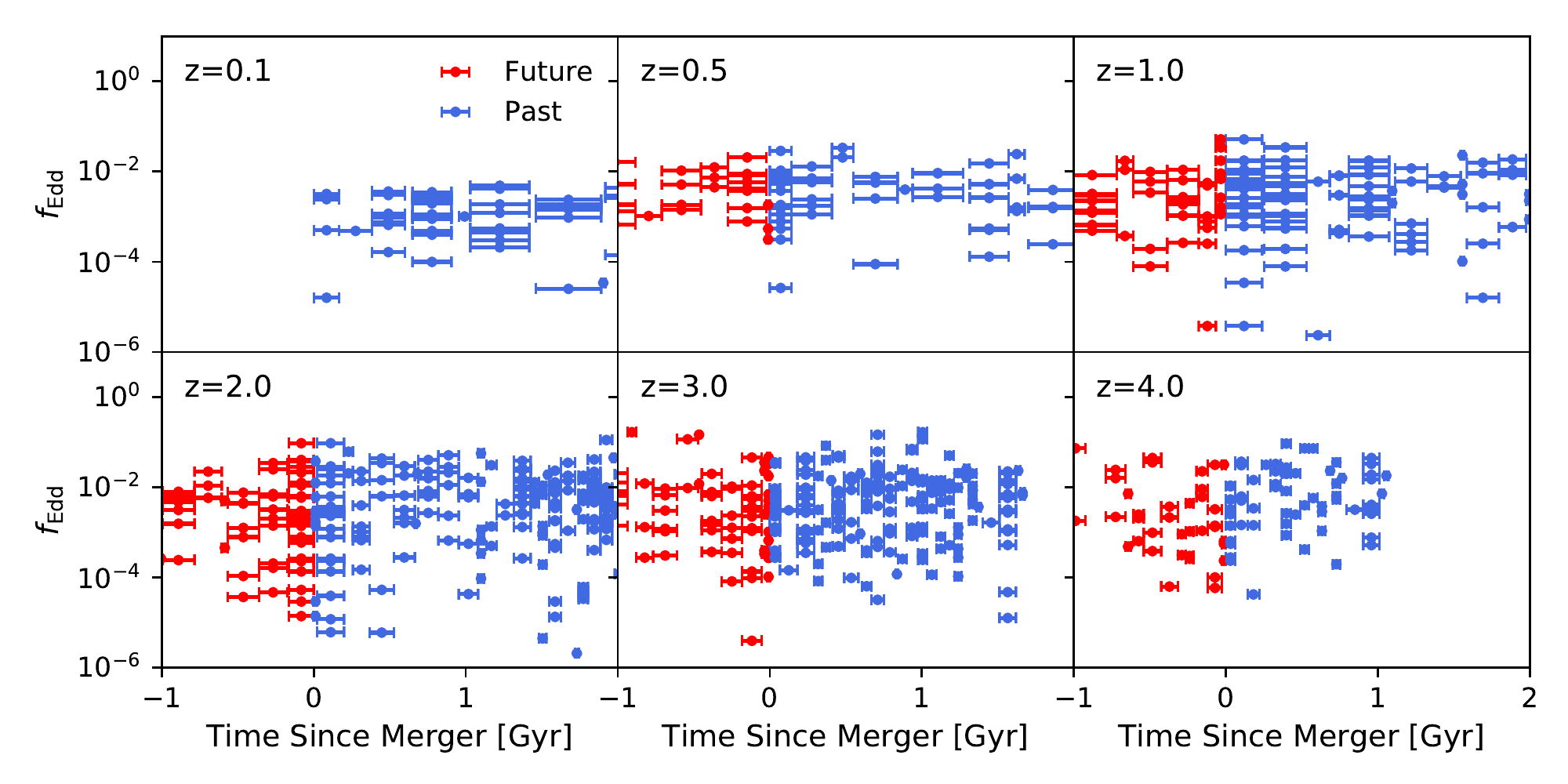}
   \caption{Eddington ratios versus time since a merger ({\it blue}) or time to a merger ({\it red}) of mass ratio $>$1:4 in {\sc Romulus25}.  No trend is evident at any timescale or any redshift.
   \label{fig:mergers}}
\end{figure*}

To investigate the AGN-merger connection, we compute merger histories for all galaxies in {\sc Romulus25} at $z=0.05$ which have a mass of at least $10^8 \ \rm{M}_\odot$, following the main progenitor branch.  Then, for different redshift slices, we compute for each progenitor (i) its Eddington ratio, (ii) the time since its most recent major merger, and (iii) the time to its next major merger.  Here, a major merger is defined as a stellar mass ratio of at least 1:4, and we find that our results do not change if this ratio is lowered to 1:10.  Although we claim in \S\ref{sec:erdfs} that the Eddington limit does not play a large role for SMBHs in {\sc Romulus}, we employ it in this section in order to treat SMBHs of all masses equally.  To help mitigate AGN variability, we convolve all Eddington ratio time series with a Gaussian kernel with a standard deviation of 30 Myr.

We plot the dependence of Eddington ratio on time since major merger in Fig. \ref{fig:mergers}.  Horizontal error bars indicate the spacing between saved simulations snapshots, since mergers can only be detected in between them.  There is no evidence for a trend between the time since merger and Eddington ratio, regardless of the time-scale or the redshift.  We also plot future mergers, in case there is a delay between merger-triggering and AGN activity.  In particular, idealized galaxy merger simulations have found that AGN should be triggered at second pericentre passage, which may precede the point at which haloes are considered merged by the halo finder \citep{Volonteri+2015b,Capelo+2015}.  We similarly find no correlation with Eddington ratio and the time to future merger.  

Our results diminish the role of the intergalactic environment in determining the SMBH accretion rate.  That said, the limited resolution of these simulations, both spatially and temporally, may hamper our ability to detect a small-scale AGN-merger connection which might be critical to the fuelling problem.  For example, our simulations would not be able to detect an AGN-merger connection if in reality it is driven by angular momentum losses on scales $\lesssim 250$ pc, or if AGN rely on large quantities of cold, molecular gas which we do not track.  We may also miss very brief bursts of merger-triggered AGN activity, which could be selected in a flux-limited survey.  Finally, as mentioned in \S\ref{sec:accretion_density}, {\sc Romulus25} lacks the volume to test the hypothesis that only the most luminous AGN are merger-triggered.  Ultra-luminous Infrared Galaxies (ULIRGs) have $z\sim 2$ number densities of $\sim 10^{-5} \ \mathrm{Mpc}^{-3}$, for example \citep{Magnelli+2011}, the kind of abundance that is inaccessible in these simulations.

\section{Discussion}\label{sec:discussion}

\subsection{SMBH-Galaxy Co-evolution}\label{sec:coevolution}

\begin{figure*}
   \centering
   \includegraphics[width=\textwidth]{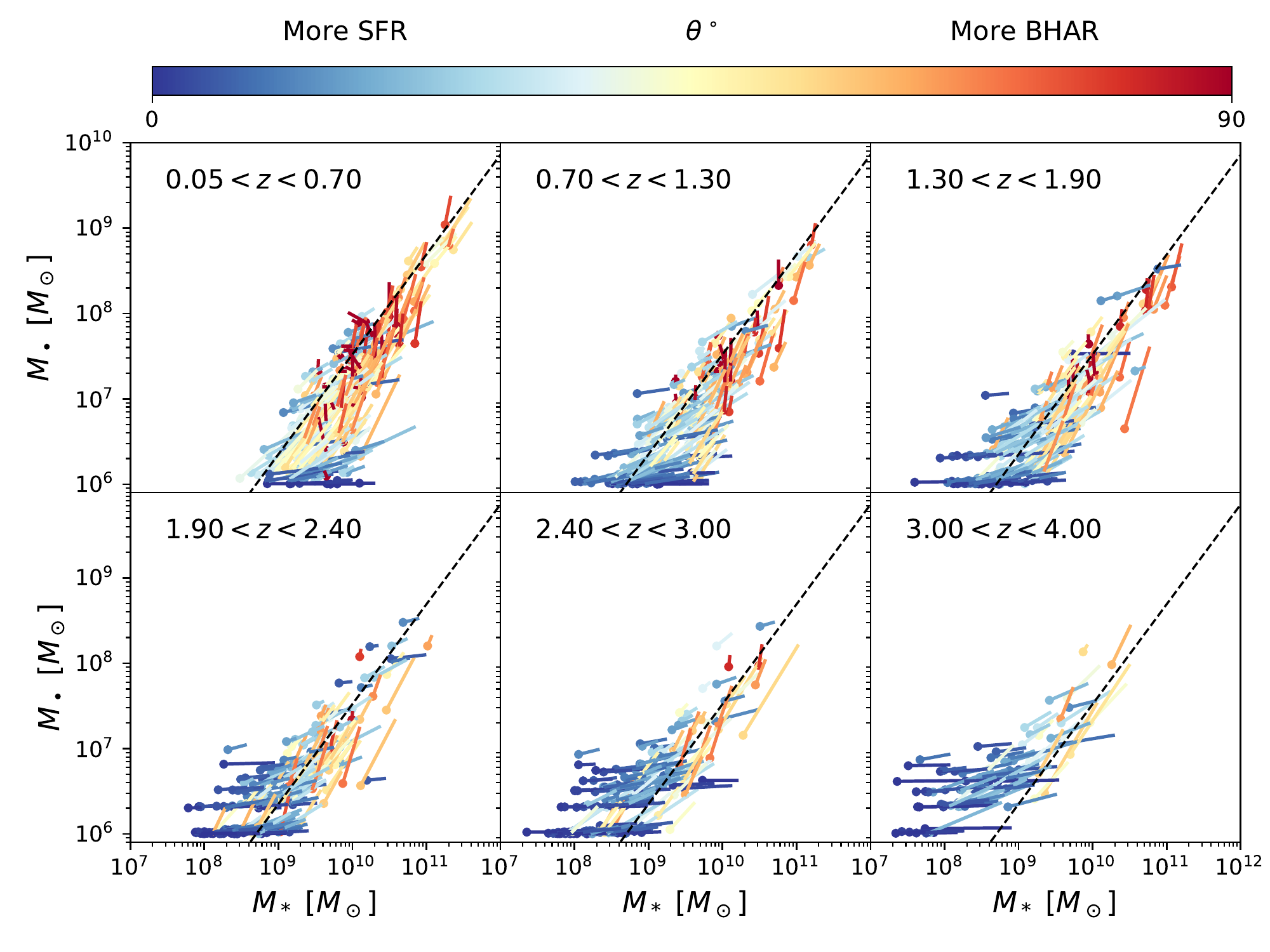}
   \caption{SMBH-Galaxy co-evolution out to $z=4$.  The evolution of galaxies that have stellar masses above $5\times 10^9 \ M_\odot$ at $z=0$ are shown.  The dashed black line displays the local $M_\bullet-M_*$ relation reported by \citet{Kormendy&Ho2013}.  Line segments are colour-coded according to the angle $\theta = \arctan(\Delta \log M_\bullet/ \Delta \log M_*)$.  SMBHs and their hosts co-evolve, attracted towards a line of constant $M_\bullet/M_*$.\label{fig:coevolution}}
\end{figure*}

The existence of a universal correlation between SFR and BHAR in the {\sc Romulus} suite implies that SMBHs and their hosts tend to grow in tandem.  The evolution of galaxies in {\sc Romulus25} on the $M_\bullet-M_*$ plane is illustrated in Figure \ref{fig:coevolution}.  Here, we begin with every galaxy at $z=0.05$ that has a stellar mass of at least $5 \times 10^9 \ \mathrm{M}_\odot$.  For each galaxy, we track back the mass of its most massive SMBH and its stellar mass from one redshift to another, and plot the resulting line segment.  For ease of readability, each segment is colour-coded according to the angle it makes with the x-axis ($\tan \theta = \Delta \log M_\bullet / \Delta \log M_*$).  The dashed black like corresponds to the relation reported by \citet{Kormendy&Ho2013}.  

SMBHs which start off overmassive with respect to their hosts (due to the high seed mass) at high-redshift do not overgrow.  As we have shown in \S\ref{sec:agnms}, their growth is capped according to their host's SFR, even if this results in low Eddington ratios.  Note that a constant $\Delta M_\bullet/\Delta M_*$ does not imply a constant $\Delta \log  M_\bullet/\Delta \log M_*$.  Instead, a SMBH which is overmassive with respect to its host will grow at a lower Eddington ratio than one which is undermassive.  This is because SMBH growth is limited by their host's gas supply rather than by its own mass \citep{Pacucci+2017}.  The result is that SMBHs and their hosts are attracted towards a line of constant SMBH-to-stellar mass ratio, moving along this line once they have reached it.  Self-regulation via feedback might be the source for this confinement \citep{Natarajan&Treister2009}.  At the highest stellar masses, the SMBH growth may proceed in galaxies which have quenched their star formation.  In addition, stripping causes some galaxies to exhibit backwards-facing gradients.

At low-redshift, there are galaxies with low-mass SMBHs whose growth does not keep up with the SFR.  This population is reflected in the low-$M_\bullet$ scatter in Fig. \ref{fig:mass_relations} as well as the low-$M_\bullet/M_*$ galaxies in Fig. \ref{fig:agnms_R25}.  We will investigate the properties of these galaxies in future work.

\subsection{Implications for Future Modelling}

The {\sc Romulus} simulations indicate that regardless of stellar mass, redshift, or intergalactic environment, the BHAR follows the SFR for star-forming galaxies.  If the SMBH-to-stellar mass ratio and the cold gas fraction are included, the $R^2$ values of these models in Table \ref{tab:AGNMS} indicate that up to 66 or 68 per cent of the variations in the BHAR can be explained.  However, while these correlations are significant, the underlying causal connection with these additional variables is still unclear.  We have shown that AGN variability significantly increases the scatter about the mean relation, even on timescales between 1.6 and 300 Myr.  AGN ``flicker'' on yet shorter timescales may continue to increase the scatter.

At the same time, it is insufficient to universally prescribe SMBH growth to equal some fraction of the SFR plus some independent scatter.  This has often been the assumption in several early semi-analytic models that trace black hole mass assembly over cosmic time.  {\sc Romulus} provides two similar lines of evidence against such a simple picture.  First, the $M_\bullet-M_*$ relation at $z=0$ has substantial scatter at low-masses, while under the simple picture all of the scatter would average out.  Such SMBH accretion prescriptions were explored in the semi-analytic model of \citet{Ricarte&Natarajan2018a}, which indeed yielded little scatter in the $M_\bullet-\sigma$ relation at low-masses.  Second, the BHAR is shown to have a positive dependence on the SMBH-to-galaxy mass ratio (see \S\ref{sec:statistical_search}).  This means that the BHAR is auto-correlated on timescales of at least 300 Myr.  A SMBH which has grown efficiently in the past is more likely to continue to do so.  

SMBHs in {\sc Romulus} can grow efficiently in hosts of any stellar mass $>10^8 \ \mathrm{M}_\odot$.  This is in contrast with many other similar cosmological simulations in which the growth of SMBHs is suppressed by supernova feedback.  These differences may be due to any combination of resolution, accretion prescriptions, supernova feedback prescriptions, and AGN feedback prescriptions.  One of the key differences in the {\sc Romulus} simulations is a unique fuelling prescription in which rotational support suppresses the BHAR.  We speculate that if for some reason the rotational support is higher in high-mass galaxies than in low-mass ones, recalibrating to the local $M_\bullet-M_*$ relations could promote more SMBH growth in low-mass galaxies.  In addition, {\sc Romulus} only contains one mode of AGN feedback (thermal), which may again alter calibrations.  There is no observational evidence for a break in the $M_\bullet-M_*$ relation down to stellar masses of $10^9 \ \rm{M}_\odot$ from broad-line relations, although this area of parameter space is sparsely sampled observationally at the present time \citep{Reines&Volonteri2015}.  It is not clear whether {\sc Romulus} or other simulations better represent the growth of BHs in low-mass galaxies.

We find no link between the BHAR and galaxy mergers, joining a body of other state-of-the-art cosmological simulations which have called into question their role in triggering AGN \citep{McAlpine+2018,Steinborn+2018,Martin+2018}.  This work favours a picture in which secular processes steadily assemble both stars and SMBHs until the galaxy is quenched.  Nevertheless, we caution that cosmological simulations currently lack the resolution to definitively settle this question.  High-resolution studies reveal that the assembly histories of SMBHs can change dramatically as resolution is increased \citep{Angles-Alcazar+2017}.  It is possible, for example, that AGN are only triggered by extreme gas densities that cannot be resolved in current cosmological simulations that can only be feasibly provided by galaxy mergers.  In addition, we cannot rule out the possibility that the rare, luminous quasars or the growth of the most massive SMBHs are triggered by mergers.  Such rare objects occur too infrequently to be found in the volume of {\sc Romulus25}, motivating future high-redshift zoom-in simulations to test this picture in more extreme environments. Although more galaxies are quenched in {\sc RomulusC} than in {\sc Romulus25}, the same AGN Main Sequence appears in both environments.  This suggests that the typical SMBH does not care about the intergalactic environment and is only affected by physical processes that are internal to the galaxy.  

\section{Conclusion}\label{sec:conclusion}

{\sc Romulus25} and {\sc RomulusC} are state-of-the-art cosmological simulations that deploy novel models for SMBH seeding, accretion, and dynamics, making them ideal for probing the drivers of SMBH growth in realistic environments.  The main results of our study are as follows:
\begin{itemize}
    \item In star-forming galaxies, the BHAR follows the SFR.  There is no dependence on redshift, stellar mass, or even large-scale environment.
    \item The SMBH-to-stellar mass ratio and cold gas fraction are shown statistically to be secondary and tertiary parameters in determining the accretion rate for star-forming galaxies.  By including these extra parameters, and by averaging over 300 Myr, up to 68 per cent of the variations in the BHAR can be explained.  However, AGN variability on shorter timescales reduces instantaneously observable correlations.
    \item SMBHs in {\sc Romulus} grow efficiently even in low-mass galaxies, in contrast with other similar cosmological simulations.
    \item The {\sc Romulus} simulations do not exhibit any link between galaxy mergers and the SMBH accretion rate for the typical galaxy, although we cannot rule out a connection for rare transient objects such as ULIRGS or $z=6$ quasars, which do not occur in the simulation's volume.
    \item AGN are variable enough on timescales between 10 and 100 Myr to complicate studies of the link between the BHAR and local conditions in the host galaxy.
\end{itemize}

These results suggest that SMBHs simply consume a fraction of the gas which forms stars, regardless of large-scale environment or even the host's stellar mass.  One major takeaway for demographic modelling is that setting the SMBH accretion rate to be proportional to the total SFR appears to be very well supported.  Crucially, however, the degree to which SMBH accretion rates are above or below the average relationship is auto-correlated in time.  The significant scatter in the $M_\bullet-M_*$ relationship at $z=0$ implies that the scatter between BHAR and SFR does not simply average out with time.  Rather, some as of yet unknown parameters favour the growth of SMBHs in some galaxies over others.  In future work, we will further investigate the scatter about the mean relation:  what drives some SMBHs to accrete more or less efficiently than the mean.

\section*{acknowledgements}  We thank our referee, Kastytis Zubovas, for his insightful comments that substantially improved the content of this paper. We thank Daisuke Nagai, Nico Cappelluti, Daniel Angles-Alcazar and Mila Chadayammuri for illuminating discussions.  We thank Tonima Tasnim Ananna for providing the luminosity densities from her model.  AR is supported by NASA Headquarters under the NASA Earth and Space Science Fellowship Program - Grant 80NSSC17K0459. MT gratefully acknowledges support from the YCAA Prize Postdoctoral Fellowship.  TQ and MT were partially supported by NSF
award AST-1514868. This research is part of the Blue Waters sustained petascale computing project, which is supported by the National Science Foundation (awards OCI-0725070 and ACI1238993) and the state of Illinois. Blue Waters is a joint effort of the University of Illinois at Urbana-Champaign and
its National Center for Supercomputing Applications. This
work is also part of a Petascale Computing Resource Allocations allocation support by the National Science Foundation (award number OAC-1613674). This work also used the
Extreme Science and Engineering Discovery Environment
(XSEDE), which is supported by National Science Foundation grant number ACI-1548562. Resources supporting this
work were also provided by the NASA High-End Computing
(HEC) Program through the NASA Advanced Supercomputing (NAS) Division at Ames Research Center.

\bibliography{ms}

\end{document}